\documentclass[showpacs,preprintnumbers,amsmath,amssymb, nofootinbib]{revtex4}

\providecommand{\abs}[1]{\lvert#1\rvert}
\usepackage{fontenc}
\usepackage{graphicx}

\usepackage{amsmath}

\begin{document}

\title{Quasinormal modes of  three-dimensional rotating
Ho\v{r}ava AdS black hole and the approach to thermal equilibrium
}

\author{Ram\'{o}n B\'{e}car}
\email{rbecar@uct.cl}
\affiliation{Departamento de Ciencias Matem\'{a}ticas y F\'{\i}sicas, Universidad Cat\'{o}lica de Temuco, Montt 56, Casilla 15-D, Temuco, Chile}
\author{P. A. Gonz\'{a}lez}
\email{pablo.gonzalez@udp.cl}
\affiliation{Facultad de Ingenier\'{\i}a y Ciencias, Universidad Diego Portales, Avenida Ej\'{e}%
rcito Libertador 441, Casilla 298-V, Santiago, Chile.}
\author{Eleftherios Papantonopoulos}
\email{lpapa@central.ntua.gr}
\affiliation{Department of Physics, National Technical University of Athens, Zografou Campus GR 157 73, Athens, Greece.}
\author{Yerko V\'{a}squez}
\email{yvasquez@userena.cl}
\affiliation{Departamento de F\'{\i}sica y Astronom\'{\i}a, Facultad de Ciencias, Universidad de La Serena,\\
Avenida Cisternas 1200, La Serena, Chile.}

\date{\today}
\begin{abstract}

We  compute the quasinormal modes (QNMs) of a massive scalar field in the background of a rotating three-dimensional Ho\v{r}ava AdS black hole, and we  analyze  the effect of the breaking of the Lorentz invariance on the QNMs.   Imposing on the horizon that there are only ingoing waves and  at infinity the Dirichlet boundary conditions and the
Neumann boundary condition, we calculate the oscillatory and  the decay modes of the QNMs. We find that the propagation of the scalar field is stable in this background and employing the holographic principle we find  the different times of the perturbed system to reach thermal equilibrium for the  various branches of solutions. 
\end{abstract}

\maketitle

\section{Introduction}

If  a dynamical system is  perturbed, it will return to equilibrium, and this process is completely
 determined by the poles of the retarded correlation function of the perturbation. In gravity theories,
 black holes are thermodynamical systems and perturbations of them at equilibrium are
described by the  quasi-normal modes (QNMs) \cite{Regge:1957td, Zerilli:1971wd, Zerilli:1970se, Kokkotas:1999bd, Nollert:1999ji, Berti:2009kk, Konoplya:2011qq}. The  QNMs are determined by solving the wave equation of an incident wave with the right boundary conditions.
Then the solution of the wave equation determines the complex frequencies, the real part of which gives the rate of oscillations of the wave while their complex part gives the required decay time for the system to reach
 thermal equilibrium.

The QNMs and quasi-normal frequencies (QNFs) have been the subjects of study for a long time  and have recently acquired great interest due to the detection of gravitational waves \cite{Abbott:2016blz}. Despite the detected signal being consistent with the Einstein gravity \cite{TheLIGOScientific:2016src}, there are great uncertainties in mass and angular momenta of the ringing black hole, which leaves open possibilities for alternative theories of gravity \cite{Konoplya:2016pmh} like  $f(R)$  gravity \cite{Starobinsky:1980te, DeFelice:2010aj, Nojiri:2005jg} and Galileon gravity theories \cite{Nicolis:2008in, Deffayet:2009wt,Kolyvaris:2011fk}. Also, the QNMs and the QNFs were extensively  studied in connection with the stability of black holes in Einstein gravity \cite{Wang:2000gsa, Wang:2004bv} and in modified gravity theories \cite{Konoplya:2018qov, Abdalla:2018ggo, Abdalla:2019irr}.

The gauge/gravity duality which results from the AdS/CFT correspondence \cite{Maldacena:1997re,Aharony:1999ti} stimulated the interest in calculating the QNMs and QNFs of  black holes in AdS spacetime. It was shown in \cite{Birmingham:2001pj} that  this holographic principle leads to the existence of a correspondence  between the QNMs in AdS black holes and linear response theory in scale invariant finite temperature field theory. This correspondence of the decay of perturbations in the dual conformal field theory and the QNMs in the gravity bulk was first discussed in  \cite{Horowitz:1999jd}.
Considering  the $(2+1)$-dimensional AdS black hole \cite{BTZ}, it was shown  analytically \cite{Birmingham:2001pj} that there is an   agreement between its QNFs  and the location of the poles of the retarded correlation function describing the linear response on the conformal field theory side.

In this work we will consider  a matter distribution in the background of three-dimensional rotating Ho\v{r}ava AdS black holes \cite{Sotiriou:2014gna}, parameterized by a scalar field. We will perturb the scalar field assuming that there is no back reaction on the metric. This will result in the calculation of the QNMs, which are characterized by a spectrum that  is independent of the initial conditions of the perturbation and depends only on the black hole  and probe field parameters, and on the fundamental constants of the system (for a  review see~\cite{Siopsis:2008xz}). Exact solutions for QNMs of black holes in three spacetime dimensions have been obtained in \cite{exactQNM}.

The motivation for considering the Ho\v{r}ava gravity theory is twofold. Considering  a condensed matter dynamical system,   it was argued in \cite{Nicolis:2015sra} that this condensed matter system breaks Lorentz invariance spontaneously and its excitations, the superfluid's phonons, have to non-linearly realize the spontaneously broken Lorentz boosts forcing   their interactions to have a very constrained structure. Then, to holographically describe such a system on the boundary, we need to have a Lorentz braking gravity theory in the bulk. The other motivation is, by calculating the QNMs, to see what is the effect of Lorentz breaking symmetry on the relaxation time of the dynamical system to reach thermal equilibrium on the boundary \cite{Papantonopoulos:2011zz}.

 In this context, the QNMs for four-dimensional non-reduced Einstein-aether theory was studied, and it was found in  \cite{Konoplya:2006rv} that the oscillation and damping rate of QNMs are larger than those of Schwarszschild black holes of the Einstein theory, for an effective potential that is known only numerically.
More recently, the QNMs for two kinds of aether black holes were analyzed, and it was shown that quasinormal ringing of the first kind of aether black hole is similar to that of another Lorentz violation model-the QED-extension limit of standard model. Also, it was found in \cite{Ding:2017gfw} that both the first and the second kind of aether black holes have larger damping rate and smaller real oscillation frequency of QNMs  compared to Schwarzschild black hole.

The work  is organized as follows. In Sec. \ref{Background} after giving  a brief review of the BTZ black hole, we discuss  the  three-dimensional Ho\v{r}ava gravity and its connection with three-dimensional Einstein-aether theory. In Sec. \ref{QNM} we find the QNMs analytically for massive scalar fields with circular symmetry and for a specify value of $J$. Also, for massive scalar field we show that the Klein-Gordon equation can be written as the Heun's equation, and we find the QNFs numerically by applying the pseudospectral method. Finally, our conclusions are in Sec. \ref{conclusion}.

\section{Three-dimensional Rotating Ho\v{r}ava Black Holes}
\label{Background}

In this Section, after reviewing in brief the BTZ black hole, we discuss the Ho\v{r}ava gravity and its connection with three-dimensional Einstein-aether theory.
The metric of the BTZ black hole is given by
\begin{equation}
ds^2=-\sinh^2\mu\left(r_+dt-r_-d\phi\right)^2+
d\mu^2+\cosh^2\mu\left(-r_-dt+r_+d\phi\right)^2\ .
\label{4}
\end{equation}
The angular coordinate $\phi$ has period $2\pi$, and the radii of the
inner and outer horizons are denoted by $r_-$ and $r_+$, respectively.
The dual conformal field theory on the boundary is $(1+1)$-dimensional,
the conformal symmetry being generated by two copies of
the Virasoro algebra acting separately on left- and right-moving sectors \cite{Birmingham:2001pj}.
Consequently,  the conformal field theory splits into two
independent sectors at thermal equilibrium with temperatures
\begin{equation}
T_L=(r_+-r_-)/2\pi\ , \quad T_R=(r_++r_-)/2\pi\ .
\label{5}
\end{equation}
According to the AdS${}_3$/CFT${}_2$ correspondence,
to each field of spin $s$  propagating in AdS${}_3$ there
corresponds an operator $\cal O$ in the dual conformal field theory
characterised by conformal weights $(h_L,h_R)$ with \cite{Aharony:1999ti}
\begin{equation}
h_R+h_L=\Delta \ ,\quad h_R-h_L=\pm s\ ,
\label{1}
\end{equation}
and $\Delta $ is determined in terms of the mass $m$ of the scalar field,
\begin{equation}
\Delta=1+\sqrt{1+m^2}\ .
\label{2}
\end{equation}
 For a small perturbation,
one expects that at late times the
perturbed system will approach equilibrium exponentially with a
characteristic time-scale. This time-scale is inversely proportional
to the imaginary part of the poles, in momentum space,
of the correlation function of the perturbation operator
${\cal O}$.
 For a conformal
field theory at zero temperature, the 2-point correlation functions can
be determined,
up to a normalisation, from conformal invariance.

Then two sets of poles were found \cite{Birmingham:2001pj}
\begin{eqnarray}
\omega_L&=&k-4\pi i T_L(n+h_L)  \ ,\nonumber\\
\omega_R&=&-k-4\pi i T_R(n+h_R)\ ,
\label{10}
\end{eqnarray}
where $n$ takes the integer values $(n=0,1,2,...)$.
This  set of poles characterises the decay
of the perturbation on the CFT side,
and coincides precisely with the quasi-normal frequencies
of the BTZ black hole \cite{Birmingham:2001pj}.

We now discuss the three-dimensional Ho\v{r}ava gravity, the action of which is given by \cite{Sotiriou:2011dr}
\begin{equation}
S_{H}=\frac{1}{16\pi G_{H}}\int dT d^2x N\sqrt{g}\left[L_{2}+L_{4}\right] \,,
\end{equation}
where $G_{H}$ is a coupling constant with dimensions of a length squared and the Lagrangian $L_{2}$ has the following form
\begin{equation}
 L_{2}=K_{ij} K^{ij}-\lambda K^2+\xi\left(^{(2)}R-2\Lambda\right)+\eta a_{i} a^{i}\,,
\end{equation}
where $K_{ij}$, $K$, and $^{(2)}R$ correspond to extrinsic, mean, and scalar curvature, respectively, and $a_{i}$ is a parameter related to the lapse function $N$ via $a_{i}=\partial_{i}\ln{N}$, being the line element in the preferred foliation
\begin{equation}
    ds^2=N^2dT^2-g_{ij}(dx^i+N^idT)(dx^j+N^jdT)\,.
\end{equation}
Also, $g$ is the determinant of the induced metric $g_{ij}$ on the constant-$T$ hypersurfaces. $L_4$ corresponds to a set of all the terms with four spatial derivatives that are invariant under diffeomorphisms. For $\lambda=\xi=1$ and $\eta=0$,  the action reduces to that of General Relativity. In the infrared limit of the theory the higher order terms $L_{4}$ (UV regime) can be neglected, and the theory is equivalent to a restricted version of Einstein-aether theory, through
\begin{equation}
    u_\alpha=\frac{\partial_\alpha T}{\sqrt{g^{\mu\nu}\partial_\mu T \partial \nu T}}\,,
\end{equation}
and
\begin{equation}
    \frac{G_H}{G_{ae}}=\xi=\frac{1}{1-c_{13}}\,,\, \frac{\lambda}{\xi}=1+c_2\,,\, \frac{\eta}{\xi}=c_{14}\,,
\end{equation}
where $c_{ij}=c_i+c_j$. Being the action of Einstein-aether:
\begin{equation}
S_{ae}=\frac{1}{16\pi G_{ae}}\int d^3x\sqrt{-g}(-R-2\Lambda+L_{ae})\,,
\end{equation}
where $G_{ae}$ is a coupling constant with dimensions of a length square, $g$ is the determinant of $g_{\mu\nu}$, $\Lambda$ is the cosmological constant, $R$ is the $3D$ Ricci scalar,
\begin{equation}
    L_{ae}=-M^{\alpha\beta\mu\nu}\nabla_\alpha u_\mu \nabla_\beta u_\nu\,,
\end{equation}
and
\begin{equation}
M^{\alpha\beta\mu\nu}=c_1g^{\alpha\beta}g^{\mu\nu}+c_2g^{\alpha\mu}g^{\beta\nu}+c_3 g^{\alpha\nu}g^{\beta\mu}+c_4 u^{\alpha}u^{\beta}g^{\mu\nu}\,.
\end{equation}

Another important characteristic of this theory is that only in the sector $\eta=0$, Horava gravity admits Asymptotically AdS solution \cite{Sotiriou:2014gna}. Therefore, assuming stationary and circular symmetry, the theory will admit the BTZ analogue to the three-dimensional rotating Ho\v{r}ava black holes described by metric
\begin{equation}\label{metric2}\
ds^{2}=Z(r)^2dt^{2}-\frac{1}{F(r)^2}dr^{2}-r^{2}(d\phi+\Omega(r)dt)^2~,
\end{equation}
where
\begin{equation}
F( r )^2= Z(r)^2 =-M +\frac{\bar{J}^2}{4r^2}-\bar{\Lambda}r^2~,
\end{equation}
with
\begin{equation}
\bar{J}^2=\frac{J^2+4a^2(1-\xi)}{\xi}~,~\Omega(r)=-\frac{J}{2 r^2}~,~\bar{\Lambda}=\Lambda-\frac{b^2(2\lambda -\xi-1)}{\xi}~,
\end{equation}
where $a$ and $b$ are constants that can be regarded as measures of aether misalignment, with $b$ as a measure of asymptotically misalignment, for $b\neq 0$, the aether does not align with the timelike Killing vector asymptotically. Note that when $\xi=1$ and $\lambda=1$, the solution becomes BTZ black holes, and for $\xi=1$, the solution becomes BTZ black holes with a shifted cosmological constant, $\bar{\Lambda}=\Lambda- 2b^2(\lambda-1)$. Also, $\bar{J}^2$ can be negative, when either $\xi < 0$ or $\xi > 1$, $a^2 > J^2/(4(\xi-1))$. The sign of $\bar{\Lambda}$  determine the asymptotic behavior (flat, dS, or AdS) of the metric \cite{Sotiriou:2014gna}.

In \cite{Sotiriou:2014gna} was argued that if $\xi>0$, and $\lambda>0$ the aether represents a well defined folation at large $r$ for any value of $b$. Moreover, if $\lambda \geq (1+\xi)/2$, then $\bar{\Lambda}$ is always negative for any $b$. Also, if the coupling constants are such that {$\xi > 0$, $\lambda > 1/2$} and $\lambda < (1+\xi)/2$, then $\bar{\Lambda}$√É√ë will switch sign at some value of b.

\begin{figure}[!h]
\begin{center}
\includegraphics[width=80mm]{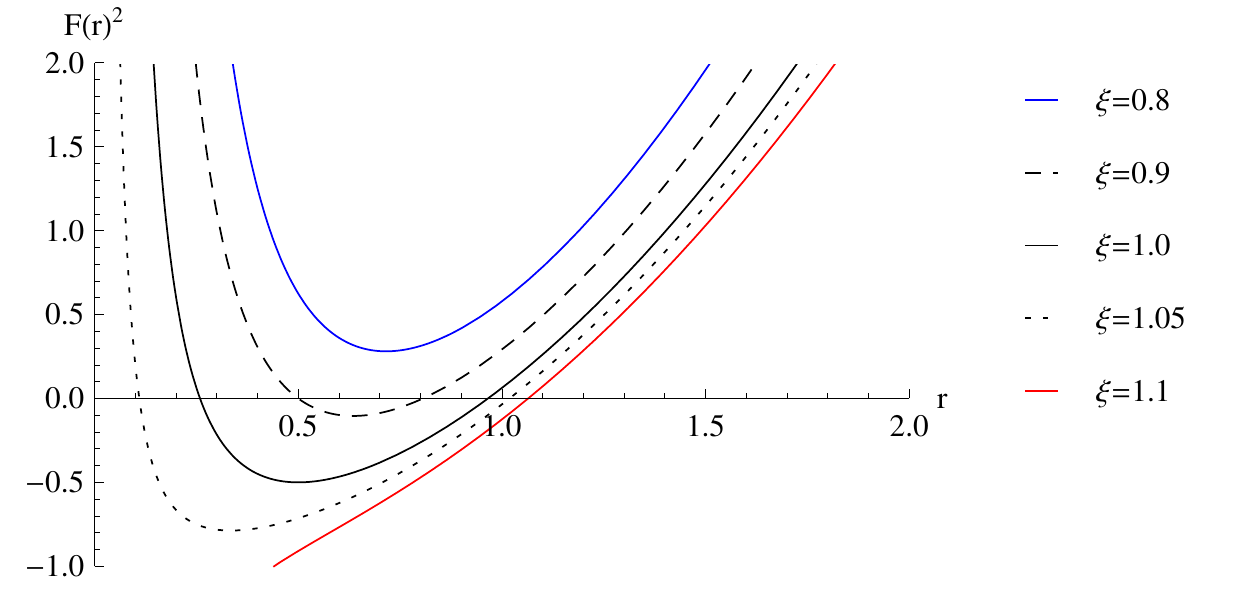}.pdf
\includegraphics[width=80mm]{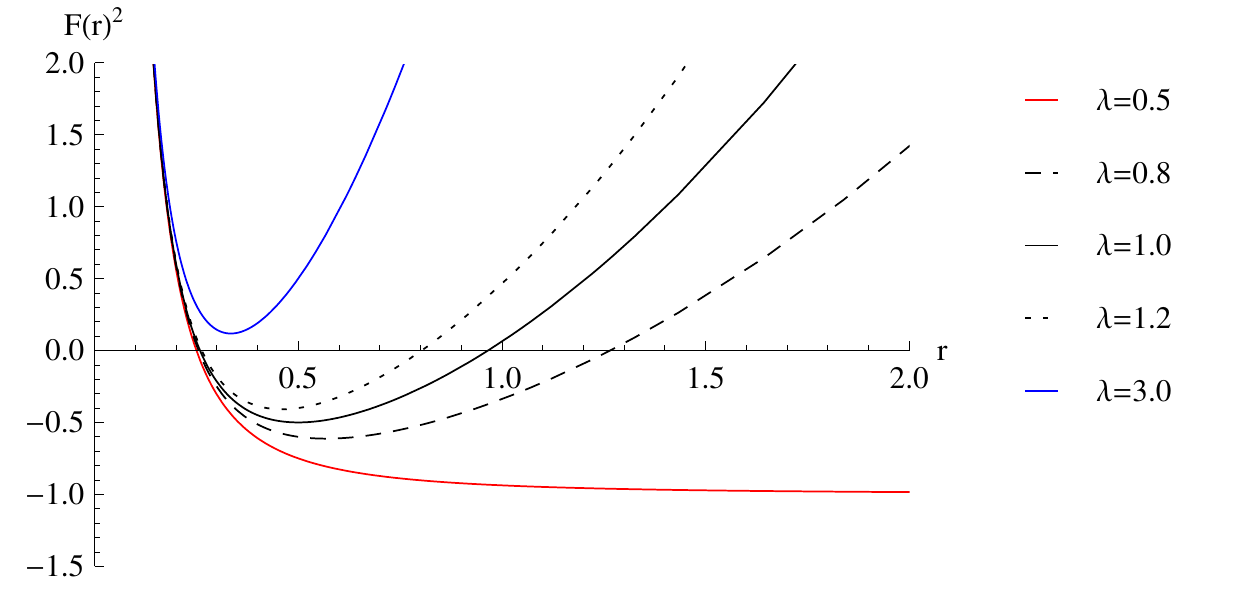}
\end{center}
\caption{The behaviour of $F(r)^2$ as a function of $r$, with $M=1$, $a=1$, $b=1$, $\Lambda=-1$, $J=0.5$. Left panel for $\lambda=1$ and right panel for $\xi=1$.}
\label{f1}
\end{figure}

\begin{figure}[!h]
\begin{center}
\includegraphics[width=50mm]{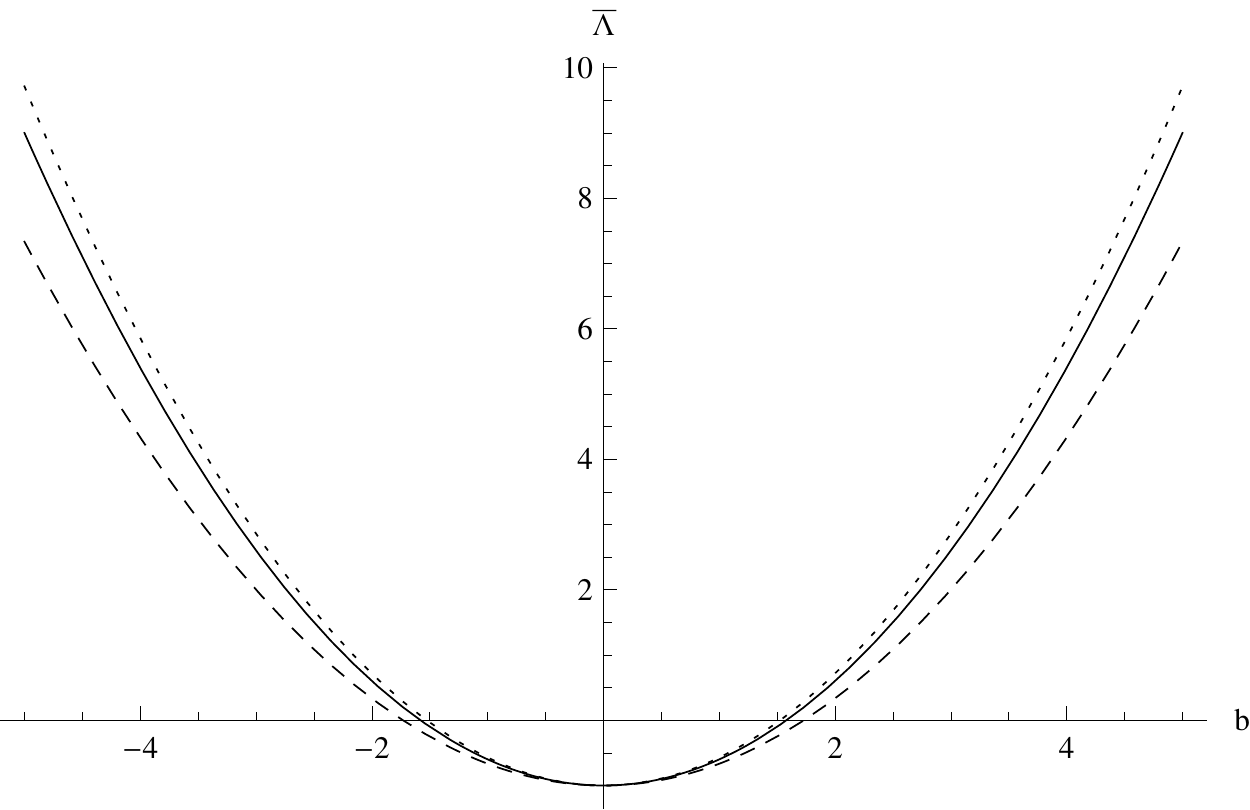}
\includegraphics[width=50mm]{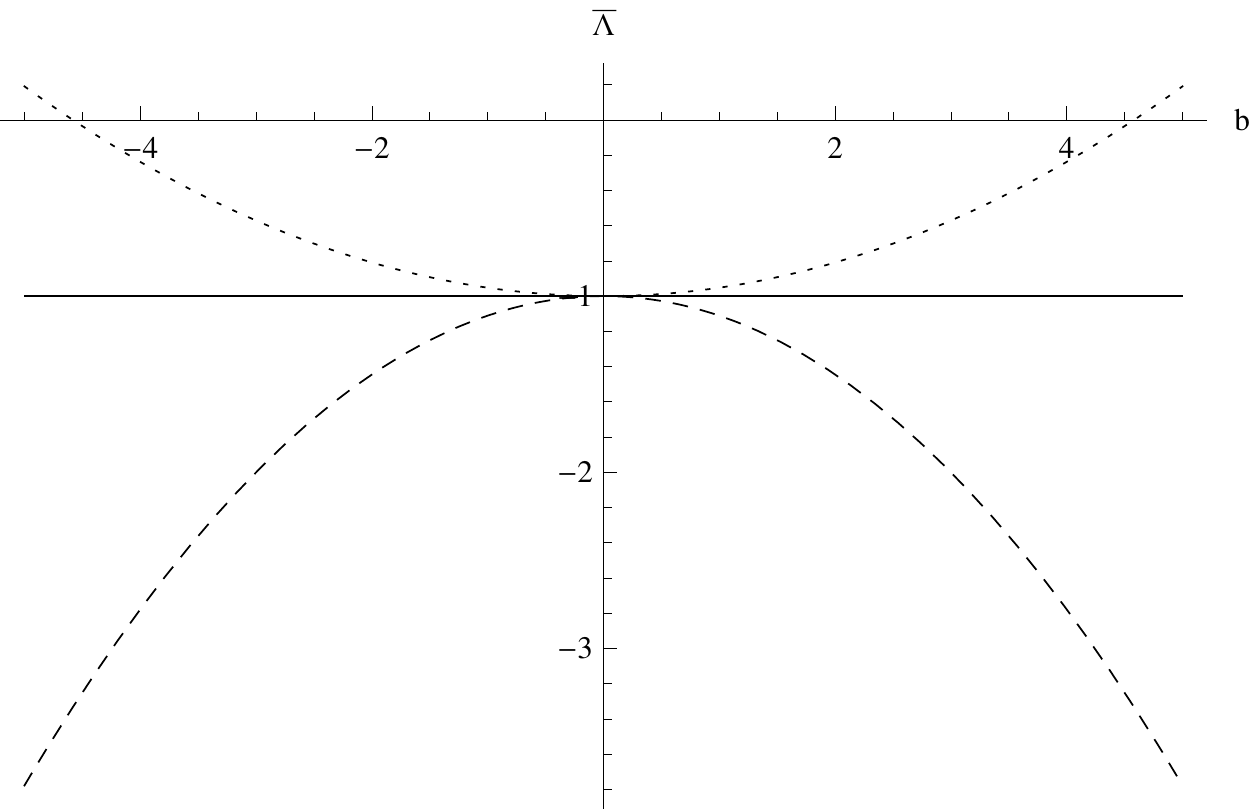}
\includegraphics[width=70mm]{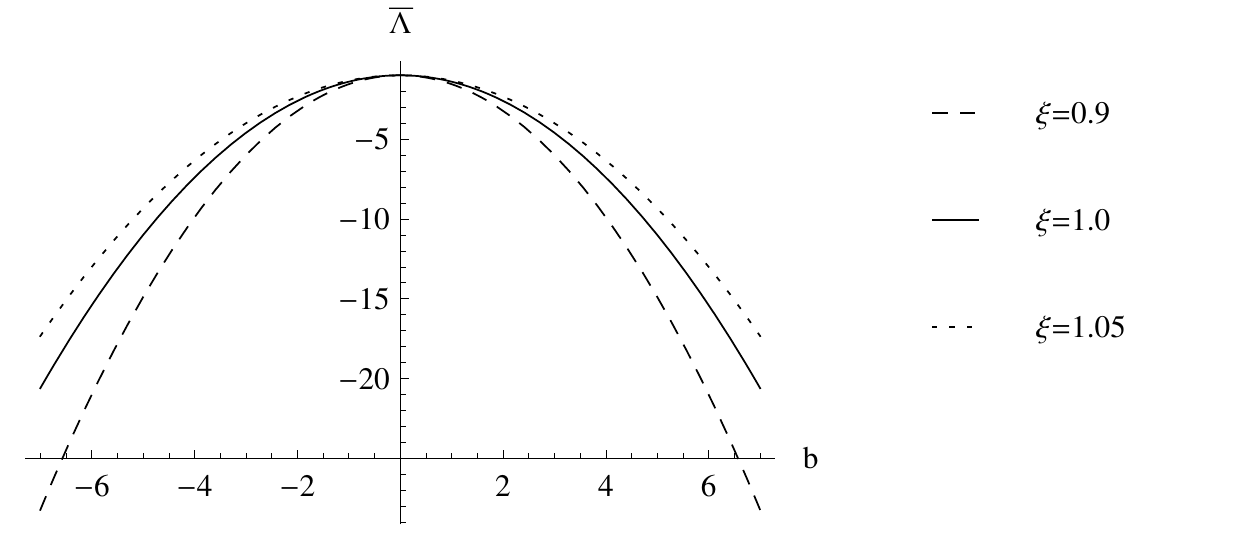}
\end{center}
\caption{The behaviour of $\bar{\Lambda}$ as a function of $b$, with $\Lambda=-1$. Left panel for $\lambda=0.8$, central panel for $\lambda=1$ and right panel for $\lambda=1.2$.}
\label{f2}
\end{figure}

\begin{figure}[!h]
\begin{center}
\includegraphics[width=50mm]{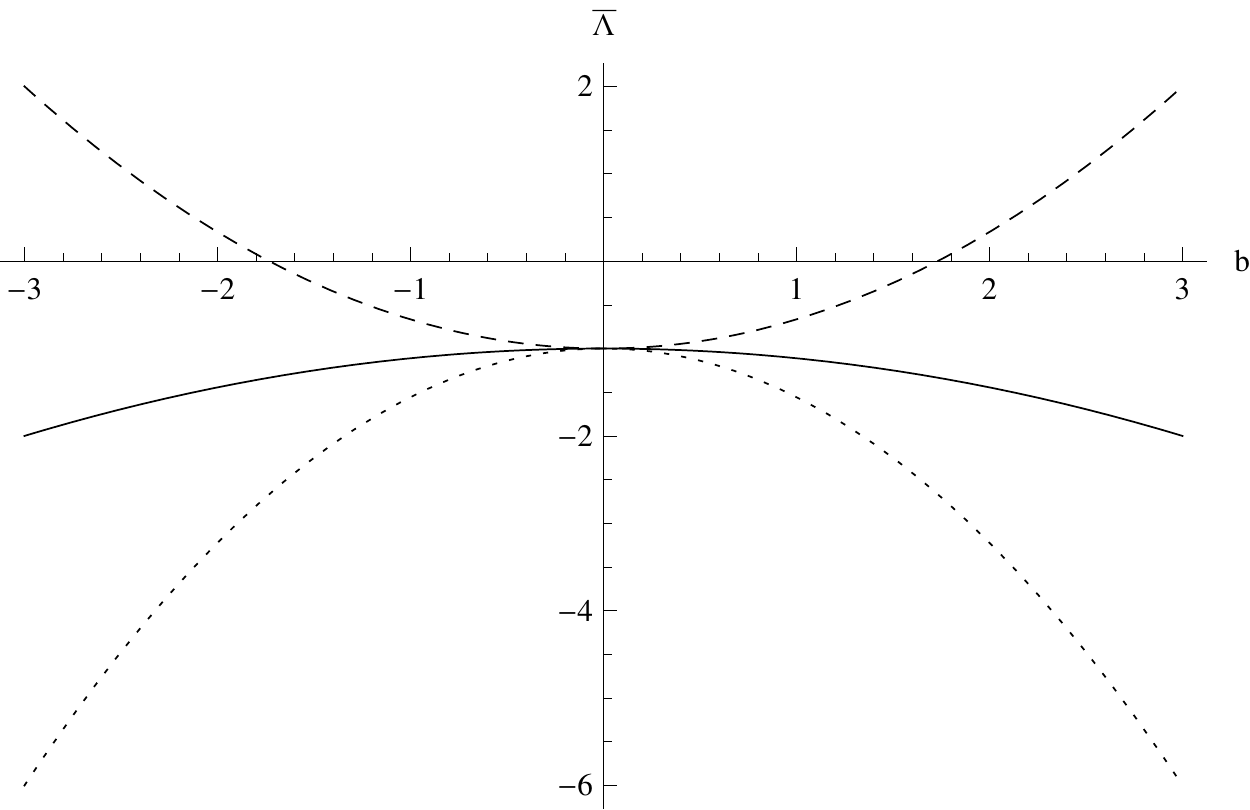}
\includegraphics[width=50mm]{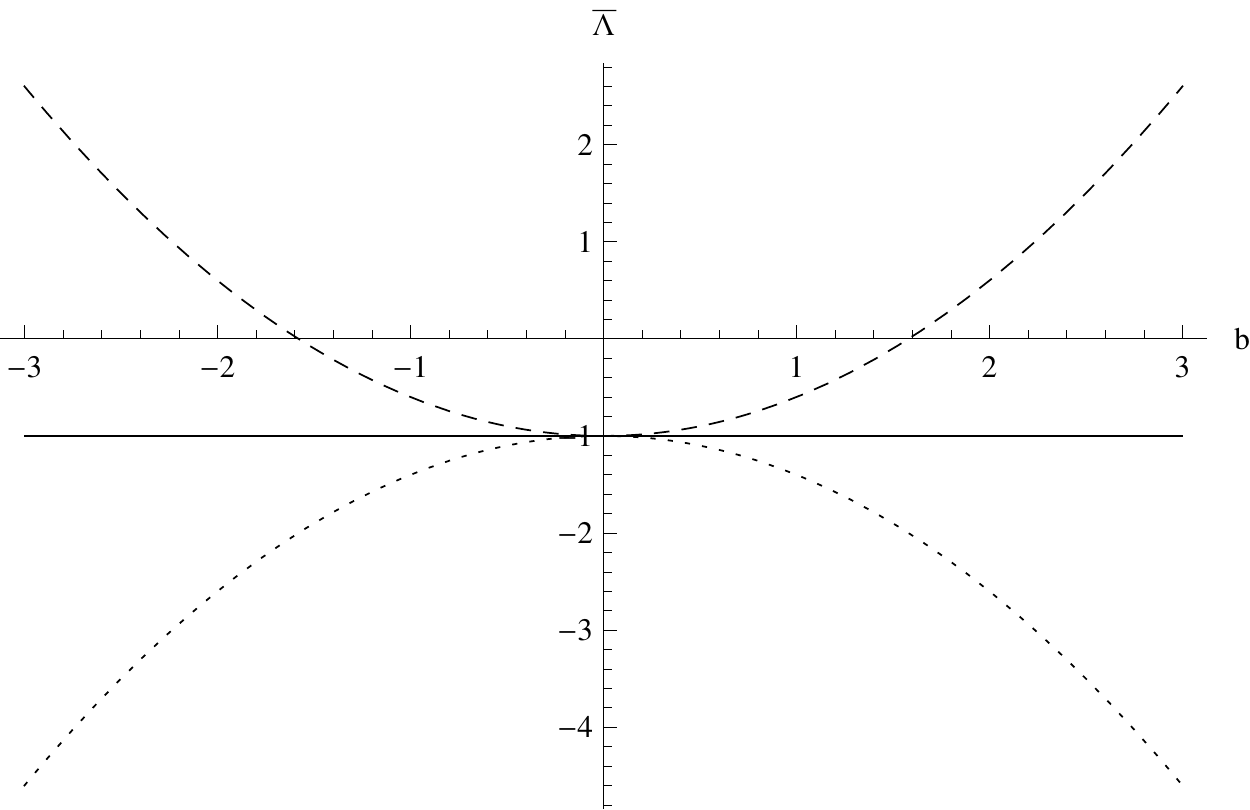}
\includegraphics[width=70mm]{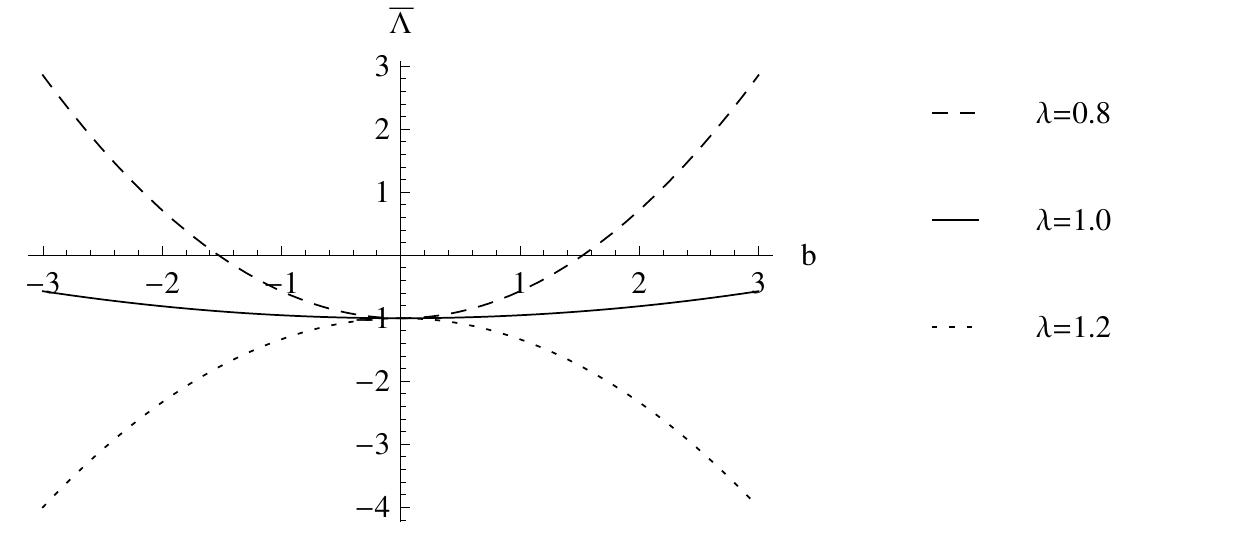}
\end{center}
\caption{The behaviour of $\bar{\Lambda}$ as a function of $b$, with $\Lambda=-1$. Left panel for $\xi=0.9$, central panel for $\xi=1$ and right panel for $\xi=1.05$.}
\label{f3}
\end{figure}

\begin{figure}[!h]
\begin{center}
\includegraphics[width=60mm]{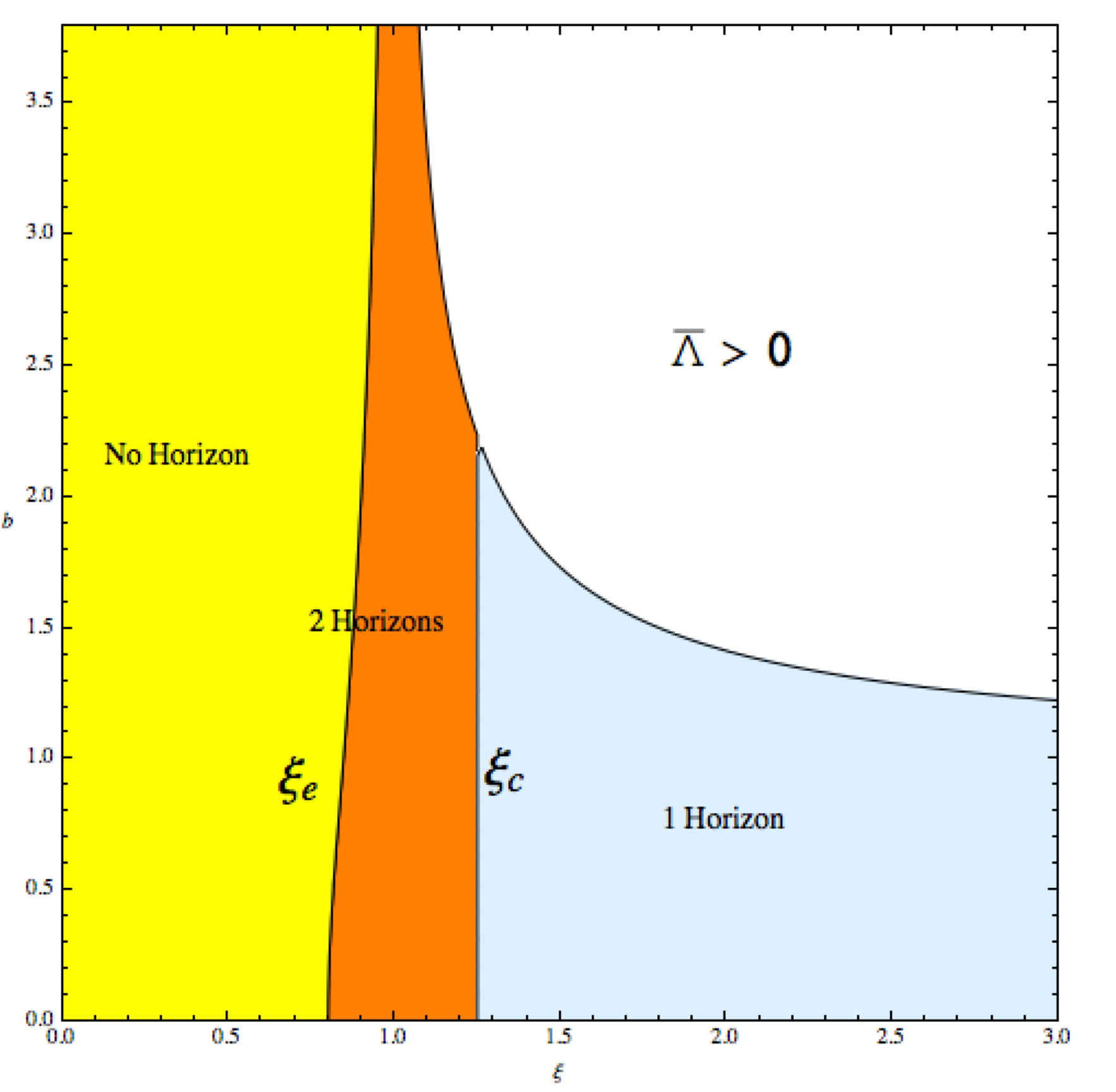}
\end{center}
\caption{Different regions of the parameter space for $a=1$, $\lambda=1$, $\Lambda=-1$, $M=1.5$ and $J=1$. The colored region corresponds to $\bar{\Lambda} <0$. In the yellow regions, there are no black hole solutions. In the orange region, the black holes have two horizons while in the light blue region, they have one horizon. }
\label{co}
\end{figure}

 In Fig. \ref{f1} we show the behavior of $F(r)^2$ as a function of $r$. When $\xi$ increases (left panel), we observe a region where there is no a horizon until ${\bf{\xi=\xi_e}}$ for which the black hole becomes extremal, this value can be obtained from $r_-=r_+$. Then there is a region where $r_+$ increases and $r_-$ decreases until becomes null when $\xi=\xi_c$, and finally there is a region $\xi>\xi_c$ where there is only one horizon. Also, we can observe the same behavior when $\lambda$ decreases (right panel). Also, in Figs. \ref{f2} and \ref{f3}, we plot the behavior of $\bar{\Lambda}$ as a function of $b$, and its  sign determines the asymptotic behavior (flat, dS, or AdS) of the metric.  Note that for $-1.4<b<1.4$, the sign of $\bar{\Lambda}$ is negative, as mentioned, if the coupling constants are such that {$\xi > 0$, $\lambda > 1/2$} and $\lambda < (1+\xi)/2$, then $\bar{\Lambda}$ will switch sign at some value of b.
The value of $\xi$ for which the black hole is extremal is given by
\begin{eqnarray}
\notag  \xi_e &=&   -\frac{1}{2(M^2-4a^2 (b^2+ \Lambda))} \Big( b^2 (J^2+8a^2 \lambda)+ \Lambda (J^2+ 4 a^2) -\Big( (b^2 (J^2+8a^2 \lambda)+ \Lambda (J^2+ 4 a^2))^2   \\
 &&  +4b^2 (J^2+4 a^2)(2 \lambda-1) (M^2-4 a^2 (b^2 + \Lambda) )\Big)^{1/2}  \Big) \, ,
\end{eqnarray}
the value of $\xi$ for which the black hole passes from having two horizons to having one horizon, an is given by
\begin{equation}
    \xi_c = \frac{4 a^2 +J^2}{4 a^2} \, ,
\end{equation}
and the value of $\xi$ for which the effective cosmological constant $\bar{\Lambda}$ changes sign is given by $\xi=\frac{(2 \lambda-1) b^2}{b^2 + \lambda}$. In Fig. (\ref{co}) we plot the different regions defined by $\xi_e$, $\xi_c$, and $\bar{\Lambda}$ for a choice of parameters.
In the case $\bar{J}\ne J$ ($\xi\ne 1$), there is a curvature singularity due to the Ricci scalar
\begin{equation}
 R=-6\bar{\Lambda}+\frac{1}{2 r^2}\left(\bar{J}^2-J^2\right)
\end{equation}
is divergent at $r=0$. This is in contrast to BTZ black holes where the Ricci and Kretschmann scalars are finite and smooth at $r=0$. The locations of the inner and outer horizons $r = r_\pm$, are given by
\begin{equation}\label{horizon}\
r_{\pm}^2=-\frac{M}{2\bar{\Lambda}}\left(1\pm \sqrt{1+\frac{\bar{J}^2\bar{\Lambda}}{M^2}} \right)~.
\end{equation}
Also, $M$ and $\bar{J}$ can be written as $M=-\bar{\Lambda}(r_+^2+r_-^2)$ and $\bar{J}= 2r_+ r_-\sqrt{-\bar{\Lambda}}$,
respectively. The Hawking temperature $T_H$ is given by
\begin{equation}
    T_H=\frac{-\bar{\Lambda}(r_+^2-r_-^2)}{2\pi r_+}\,.
\end{equation}

\section{QNMs}
\label{QNM}


The quasi-normal modes of scalar
perturbations  for a minimally coupled massive scalar field to curvature on the background of three-dimensional Ho\v{r}ava AdS black holes are described by the solution of the Klein-Gordon equation
\begin{equation}
\Box \psi = \frac{1}{\sqrt{-g}}\partial_{\mu}\left(\sqrt{-g} g^{\mu\nu}\partial_{\nu}\right)\psi=-m^2\psi~,
\end{equation}
where $m$ is the mass of the scalar field $\psi$. Which can be written as
\begin{equation}\label{de2}
\left( -\frac{1}{F(r)^2}\partial_t^2+F(r)^2\partial_r^2+\frac{1}{r}\partial_r(rF(r)^2)\partial_r-\frac{J}{r^2F(r)^2}\partial_t\partial_\phi+\frac{1}{r^2F(r)^2}(F(r)^2-\frac{J^2}{4r^2})\partial_\phi^2-m^2\right)\Psi=0~.
\end{equation}
The term $F(r)^2-\frac{J^2}{4r^2}$ is given by
\begin{eqnarray}
\nonumber -M +\frac{\bar{J}^2}{4r^2}-\bar{\Lambda}r^2-\frac{J^2}{4r^2}&=&-M + \frac{J^2+4a^2(1-\xi)}{4\xi r^2}-(\Lambda-\frac{b^2(2\lambda-\xi-1)}{\xi})r^2-\frac{J^2}{4r^2}\\
&&=-M+\frac{J^2}{4r^2}\left(\frac{1}{\xi}+ \frac{4a^2(1-\xi)}{\xi}-1\right)-\left(\Lambda-\frac{b^2(2\lambda-\xi-1)}{\xi}\right)r^2~.
\end{eqnarray}
It is worth mentioning that  the second term in the above expression vanishes for $\xi=1$. Performing the change of variables $z=\frac{r^2-r_+^2}{r^2-r_-^2}$ along with the ansatz $\psi=R(z)e^{-i\omega t}e^{i\kappa\phi}$, Eq. (\ref{de2}) yields
\begin{eqnarray}\label{radial12}\
&& z(1-z)\partial_{z}^2R(z)+\left(1-z\right)\partial_{z}R(z)+\\
\nonumber &&\left[-\frac{\omega^2(zr_-^2-r_+^2)}{4\bar{\Lambda}^2(r_+^2-r_-^2)^2z}-\frac{J\omega\kappa}{4\bar{\Lambda}^2(r_+^2-r_-^2)^2}\frac{1-z}{z}-\frac{\kappa^2}{4\bar{\Lambda}^2(r_+^2-r_-^2)^2}\left(F(z)^2-\frac{J^2}{4r(z)^2} \right)\frac{1-z}{z}+\frac{m^2}{4(1-z)\bar{\Lambda}}\right]R(z)=0~.
\end{eqnarray}

\subsection{Massive scalar field with circular symmetry}

For a massive scalar field with circular symmetry ($\kappa=0$) Eq. (\ref{radial12}) is
\begin{equation}\label{ra12}\
z(1-z)\partial_{z}^2R(z)+\left(1-z\right)\partial_{z}R(z)+
\left[-\frac{\omega^2(zr_-^2-r_+^2)}{4\bar{\Lambda}^2(r_+^2-r_-^2)^2 z}+\frac{m^2}{4(1-z)\bar{\Lambda}}\right]R(z)=0~,~
\end{equation}
which can be written as
\begin{equation}\label{r12}\
z(1-z)\partial_{z}^2R(z)+\left(1-z\right)\partial_{z}R(z)+
\left(A+\frac{B}{z}+\frac{C}{1-z}\right)R(z)=0~,
\end{equation}
where
\begin{equation}
A= -\frac{\omega^2r_-^2}{4\bar{\Lambda}^2(r_+^2-r_-^2)^2}~,~ B=\frac{\omega^2r_+^2}{4\bar{\Lambda}^2(r_+^2-r_-^2)^2} ~,~ C=\frac{m^2}{4\bar{\Lambda}}~.
\end{equation}
Under the decomposition $R(z)=z^\alpha(1-z)^\beta K(z)$, Eq.
(\ref{r12}) can be written as a hypergeometric equation for $K$
\begin{equation}\label{hypergeometric}\
z(1-z)K''(z)+\left[c_1-(1+a_1+b_1)z\right]K'(z)-a_1b_1 K(z)=0 \,,
\end{equation}
where the coefficients $a_1$, $b_1$, and $c_1$ are given by
\begin{equation}\label{a}\
a_1= \alpha+\beta\mp\sqrt{A}~,\,
b_1= \alpha+\beta\pm\sqrt{A} ~,\,
c_1=1+2\alpha~,
\end{equation}
and the exponents $\alpha$ and
$\beta$ are
\begin{equation}
\alpha=\pm i\sqrt{B}~,\,
\beta=\frac{1}{2} \left(1\pm\sqrt{1-4 C}\right) ~.
\end{equation}
The general solution of Eq. (\ref{hypergeometric}) takes
the form
\begin{equation}
K=C_{1}\,{}_2F_{1}(a_1,b_1,c_1;z)+C_2z^{1-c}\,{}_2F_{1}(a_1-c_1+1,b_1-c_1+1,2-c_1;z)~,
\end{equation}
which has three regular singular points at $z=0$, $z=1$, and
$z=\infty$. Here, ${}_2F_{1}(a_1,b_1,c_1;z)$ is a hypergeometric function
and $C_{1}$, $C_{2}$ are constants. Then, without loss of generality, we choose the negative sign for
$\alpha$, and the solution for the
radial function $R(z)$ is
\begin{equation}\label{RV}\
R(z)=C_{1}z^\alpha(1-z)^\beta
\,{}_2F_{1}(a_1,b_1,c_1;z)+C_2z^{-\alpha}(1-z)^\beta
\,{}_2F_{1}(a_1-c_1+1,b_1-c_1+1,2-c_1;z)~.
\end{equation}
According to our change of variables at the vicinity of the
horizon $r\rightarrow r_{+}$, $z\rightarrow
0$, and at infinity $r\rightarrow \infty$,
$z\rightarrow 1$. In the vicinity of the horizon, $z=0$ and using
the property $F(a_1,b_1,c_1,0)=1$, the function $R(z)$ behaves as
\begin{equation}
R(z)=C_1 e^{\alpha \ln z}+C_2 e^{-\alpha \ln z},
\end{equation}
and the scalar field $\psi$ can be written in the following way
\begin{equation}
\psi\sim C_1 e^{-i\omega(t+ \frac{r_+}{2\abs{\bar{\Lambda}}(r_+^2-r_-^2)}\ln z)}+C_2
e^{-i\omega(t- \frac{r_+}{2\abs{\bar{\Lambda}}(r_+^2-r_-^2)}\ln z)}~,
\end{equation}
in which the first term represents an ingoing wave and the second
term an outgoing wave in the black hole. To compute the QNMs, we
have to impose the boundary conditions on the horizon that  there
exist only ingoing waves. This fixes  $C_2=0$. So, the radial
solution becomes
 \begin{equation}\label{horizonsolutiond}
R(z)=C_1 e^{\alpha \ln z}(1-z)^\beta F_{1}(a_1,b_1,c_1;z)=C_1
e^{-i\omega\frac{r_+}{2\abs{\bar{\Lambda}}(r_+^2-r_-^2)}\ln z}(1-z)^\beta F_{1}(a_1,b_1,c_1;z)~.
\end{equation}
In order to implement boundary conditions at infinity ($z=1$), we
shall apply in Eq. (\ref{horizonsolutiond}), the Kummer's formula,
for the hypergeometric function \cite{M. Abramowitz},
\begin{eqnarray}
\nonumber {}_2F_{1}(a_1,b_1,c_1;z)&=&\frac{\Gamma(c_1)\Gamma(c_1-a_1-b_1)}{\Gamma(c_1-a_1)\Gamma(c_1-b_1)}
\,{}_2F_1(a_1,b_1,a_1+b_1-c_1,1-z)\\
&& +(1-z)^{c_1-a_1-b_1}\frac{\Gamma(c)\Gamma(a_1+b_1-c_1)}{\Gamma(a_1)\Gamma(b_1)}\,{}_2F_1(c_1-a_1,c_1-b_1,c_1-a_1-b_1+1,1-z)\,.
\end{eqnarray}
With this expression, the radial function results in
\begin{eqnarray}\label{R}\
 R(z) &=& C_1 e^{-i\omega\frac{r_+}{2\abs{\bar{\Lambda}}(r_+^2-r_-^2)}\ln z}(1-z)^\beta\frac{\Gamma(c_1)\Gamma(c_1-a_1-b_1)}{\Gamma(c_1-a_1)\Gamma(c_1-b_1)} \,{}_2F_1(a_1,b_1,a_1+b_1-c_1,1-z)\\
\nonumber && +C_1 e^{-i\omega\frac{r_+}{2\abs{\bar{\Lambda}}(r_+^2-r_-^2)}\ln
z}(1-z)^{c_1-a_1-b_1+\beta}\frac{\Gamma(c_1)\Gamma(a_1+b_1-c_1)}{\Gamma(a_1)\Gamma(b_1)}\,{}_2F_1(c_1-a_1,c_1-b_1,c_1-a_1-b_1+1,1-z)~.
\end{eqnarray}
Therefore, by imposing that the scalar field at infinity is null, for $m^2/\bar{\Lambda}<0$  ($\beta_-<0$ and $c_1-a_1-b_1+\beta_-=\beta_+>0$), then the term proportional to $(1-z)^\beta$ in Eq. (\ref{R}) diverges. So, we obtain that the scalar field is null only upon the following additional restriction
$(c_1-a_1)|_{\beta_-}=-n$ or $(c_1-b_1)|_{\beta_-}=-n$.  Then, the QNFs yields

\begin{equation}\label{qnmpp}
\omega_1=-i \abs{\bar{\Lambda}} \left(\sqrt{1-4C}+2n+1\right) (r_- + r_+)\,,
\end{equation}
\begin{equation}\label{qnmp2}
\omega_2=-i \abs{\bar{\Lambda}} \left(\sqrt{1-4C}+2n+1\right) (r_+ - r_-)\,,
\end{equation}
respectively. Note that the imaginary part of the QNFs is negative, which ensures that the propagation of scalar fields is stable in this background.

\begin{figure}[!h]
\begin{center}
\includegraphics[width=60mm]{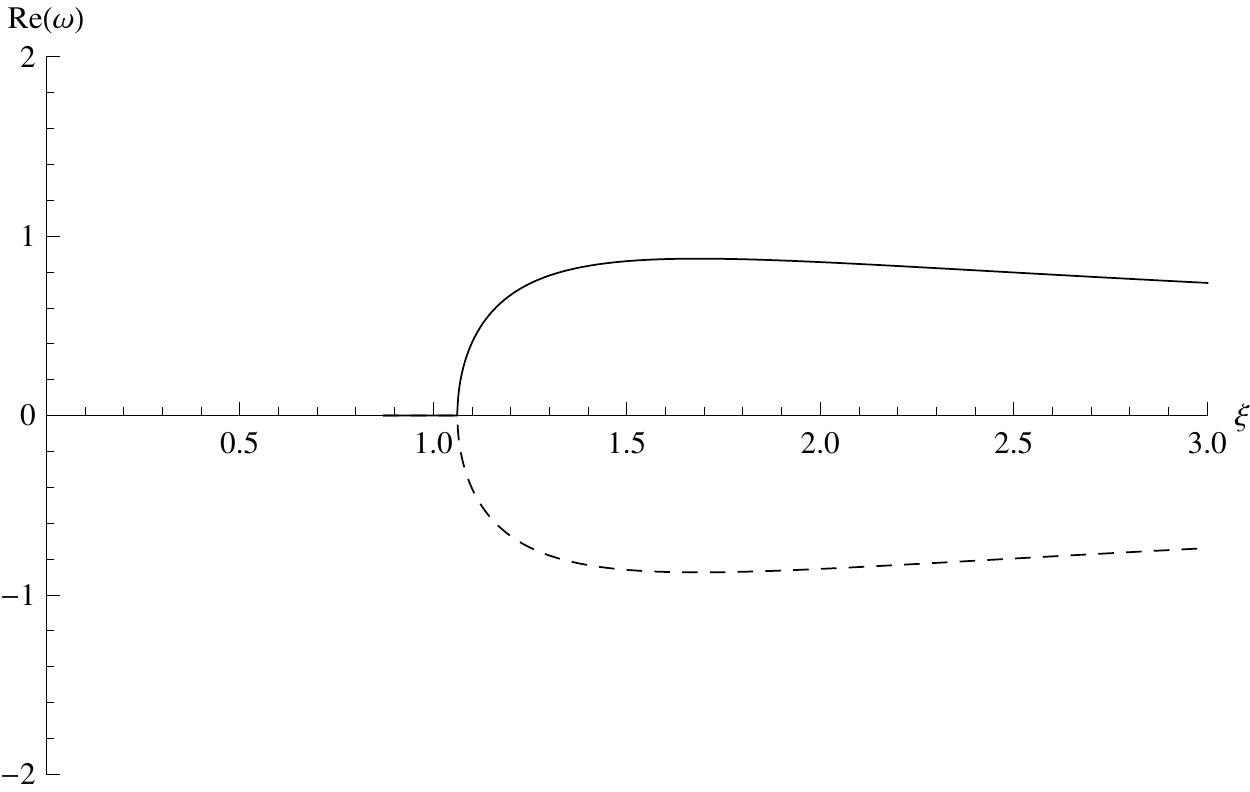}
\includegraphics[width=60mm]{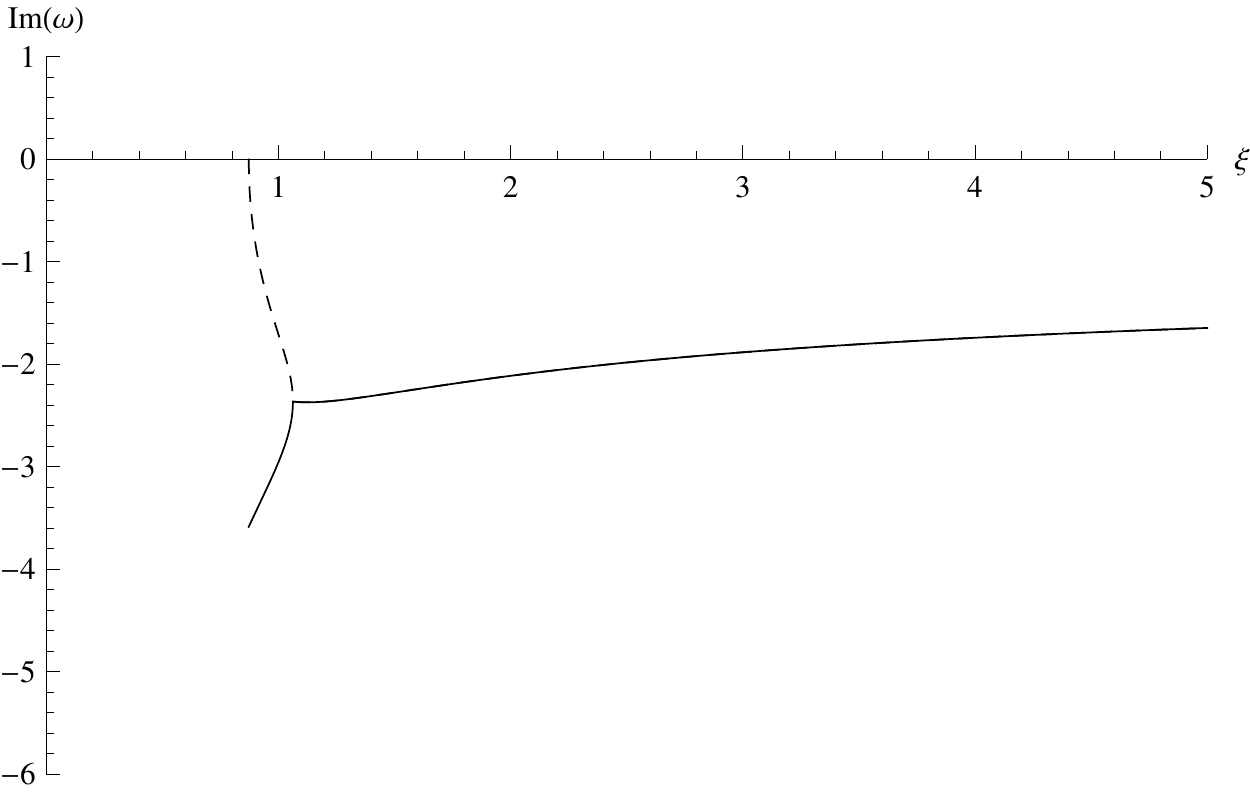}
\end{center}
\caption{The fundamental QNFs for $M=1$, $\lambda=1$, $\Lambda= -1$, $a=b=m=1$, and $J=0.5$, as a function of $\xi$. Left panel for $Re(\omega)$, and right panel for $Im(\omega)$. Continuous line for $\omega_1$, and dashed line for $\omega_2$.}
\label{omegaxi}
\end{figure}

\begin{figure}[!h]
\begin{center}
\includegraphics[width=80mm]{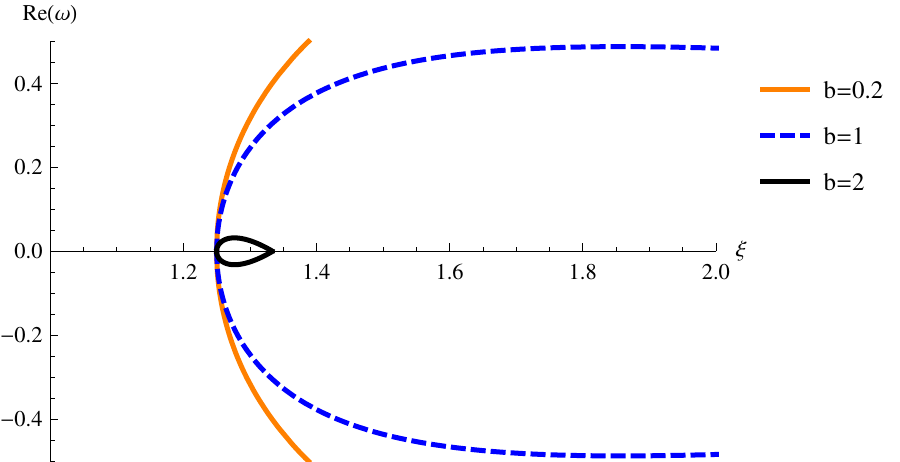}
\includegraphics[width=80mm]{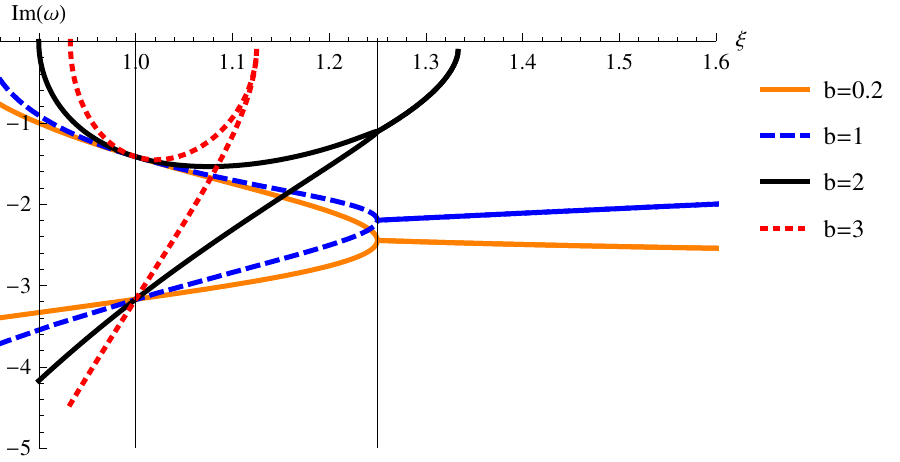}
\end{center}
\caption{The fundamental QNFs for $M=1.5$, $\lambda=1$, $\Lambda= -1$, $a=1$, $m=0.1$, $J=1$ and $b=0.2, 1,2, 3$ as a function of $\xi$. Left panel for $Re(\omega)$, and right panel for $Im(\omega)$. In the right panel, the upper curves correspond to the $\omega_1$ branch while the lower curves correspond to the $\omega_2$ branch. For $\xi=1$ all the frequencies coincide with the frequencies of BTZ black holes. For $\xi_c=1.25$, both branches converge. For $\xi < \xi_c$ the QNFs are purely imaginary while that for $\xi> \xi_c $ acquire a real part. For $b=3$, the frequencies are purely imaginary. }
\label{on}
\end{figure}

Now, in order to observe the behavior of the QNFs (\ref{qnmpp}) and (\ref{qnmp2}), we plot in Fig. \ref{omegaxi}, the behavior of the  real (left panel) and imaginary parts (right panel) of the fundamental QNFs as a function of $\xi$. Note that, as mentioned, for $\xi<\xi_e$, there is no horizon.
  So, for $\xi > \xi_e$, we observe that for $\omega_1$ (continuous line) there is a range where $Re(\omega_1)$ is null and then takes positive values, when the coupling constant $\xi$ increases, while $\abs{Im(\omega_1)}$ decreases
when $\xi$ increases. So, according to the gauge/gravity duality, the relaxation time in order to reach the thermal equilibrium increases for the right sector. However, for $\omega_2$ (dashed line) and  $\xi > \xi_e$, $Re(\omega_2)$ is null and then takes negative values, while its imaginary part increases and then decreases when the coupling constant $\xi$ increases, showing that the relaxation time can decrease or increase depending on the value of $\xi$. It is interesting to note that when $Im(\omega_2)$ decreases $Im(\omega_1)=Im(\omega_2)$. If we consider the BTZ black hole, $\xi=1$ and $\lambda=1$, the real part is null and $Im(\omega_1)\neq Im(\omega_2)$. In the following, we will analyze the two branches of QNFs for different values of $\xi$. Fig. (\ref{on}) is similar to Fig. (\ref{omegaxi}), but in order to see the effect of $b$ on the behavior of the QNFs, we have plotted several curves corresponding to different values of parameter $b$. For $\xi=1$ the QNFs coincide and correspond to the QNFs of the BTZ black hole. On the other hand, for $\xi= \xi_c$, both purely imaginary branches converge to the same value. For values of $\xi$ near 1, $|Im(\omega_1)|$ decreases when $\xi$ increases, decreasing faster for large values of $b$, while $|Im(\omega_2)|$ increases when $\xi$ increases, increasing faster for small values of $b$, which implies that the relaxation time of the right sector increases and the relaxation time of the left sector decreases. On the other hand, we observe that for $\xi$ near $\xi_c$ with $\xi> \xi_c$, only one branch exists, and $|Im(\omega)|$ decreases for small values of $b$ while it increases for large values of $b$.
Notice that for $b=3$ the effective cosmological constant $\bar{\Lambda}$ becomes positive before reaching the value $xi_c$.


\subsubsection{$\xi_e < \xi < \xi_c$}

\begin{figure}[!h]
\begin{center}
\includegraphics[width=60mm]{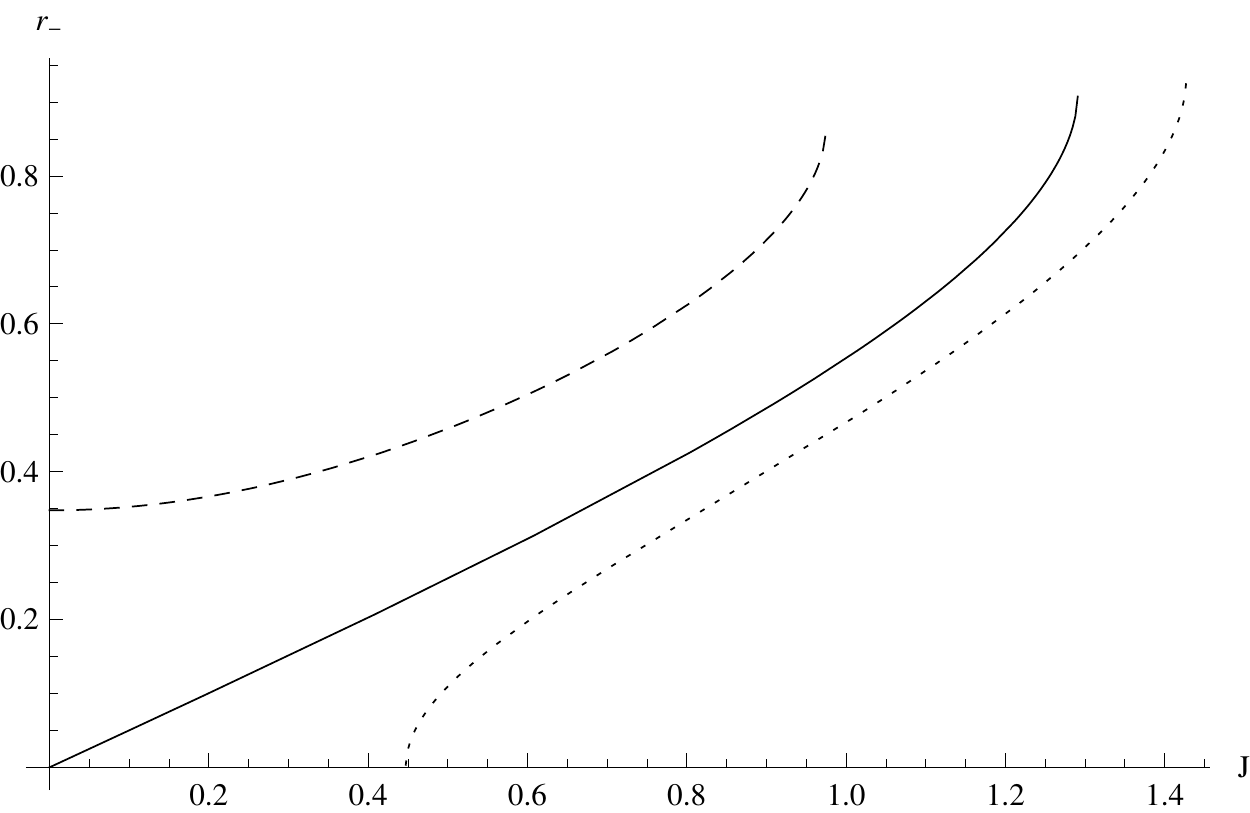}
\includegraphics[width=60mm]{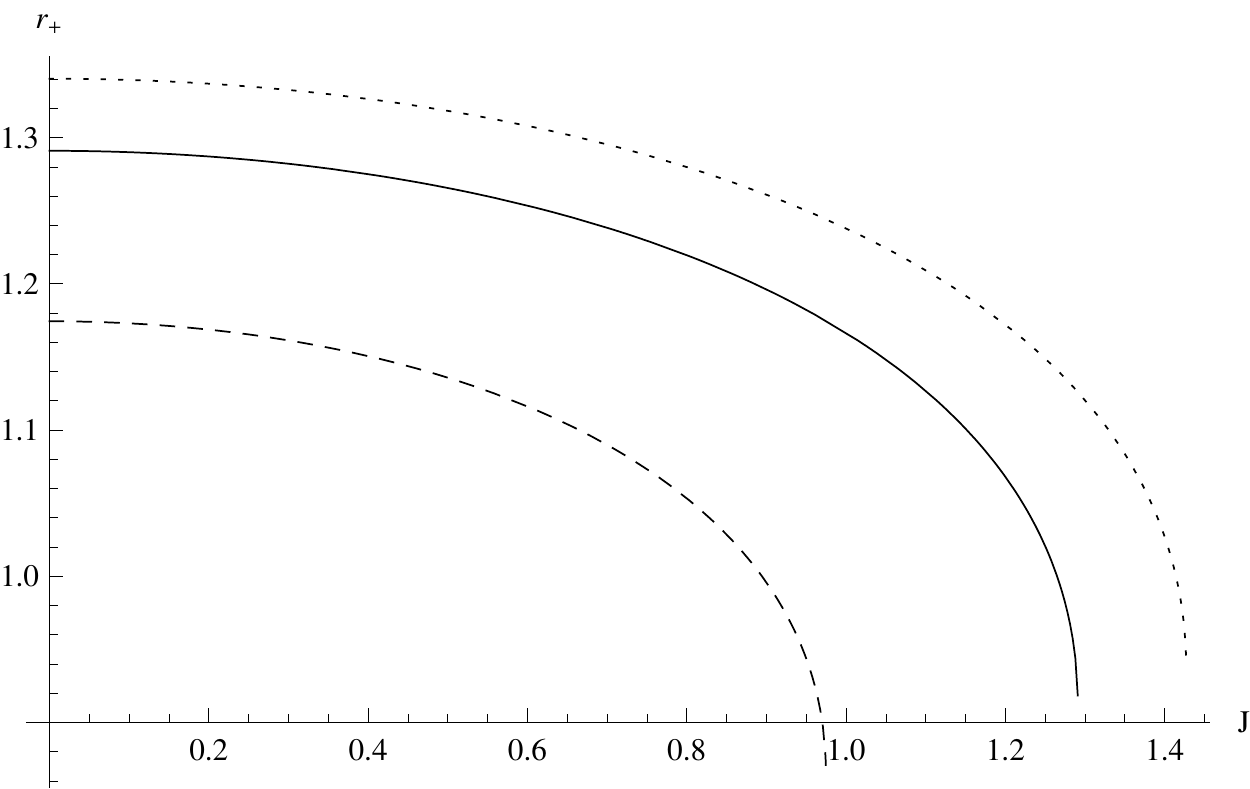}\\
\includegraphics[width=60mm]{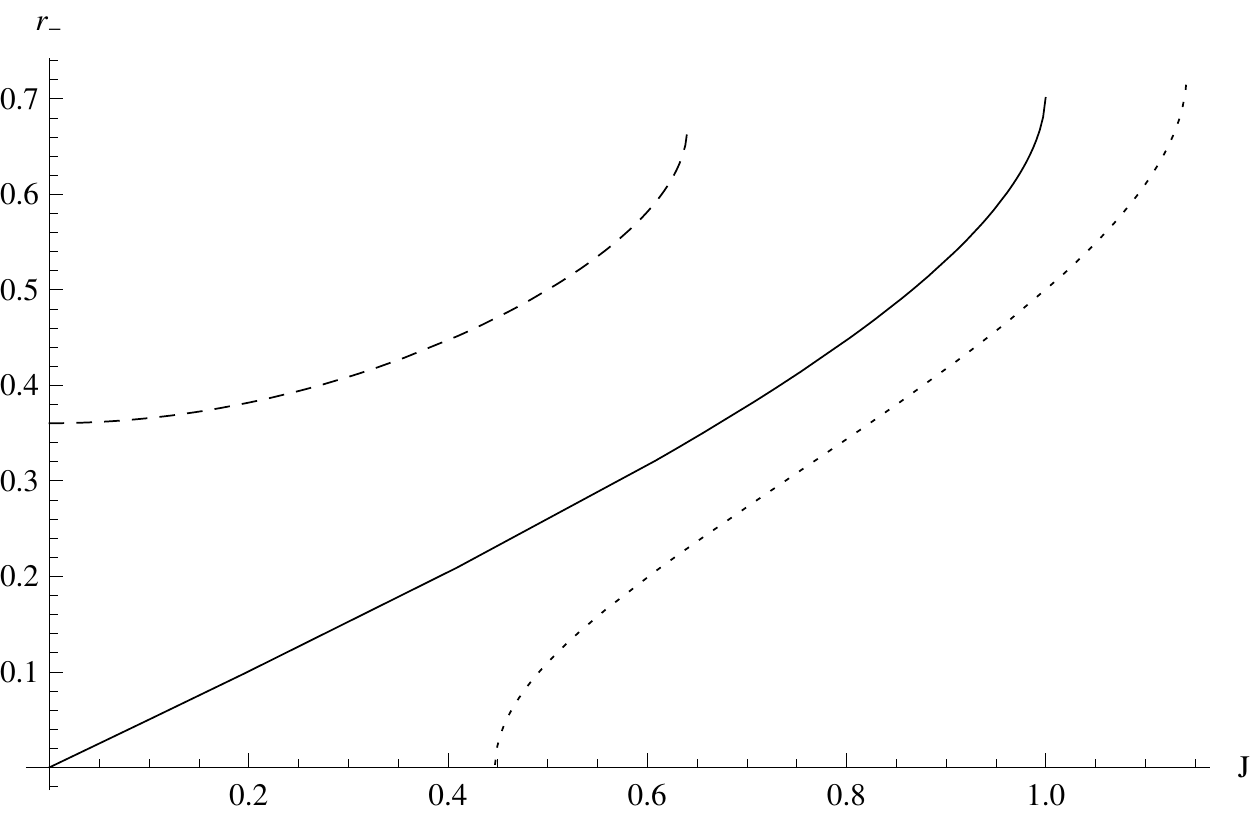}
\includegraphics[width=60mm]{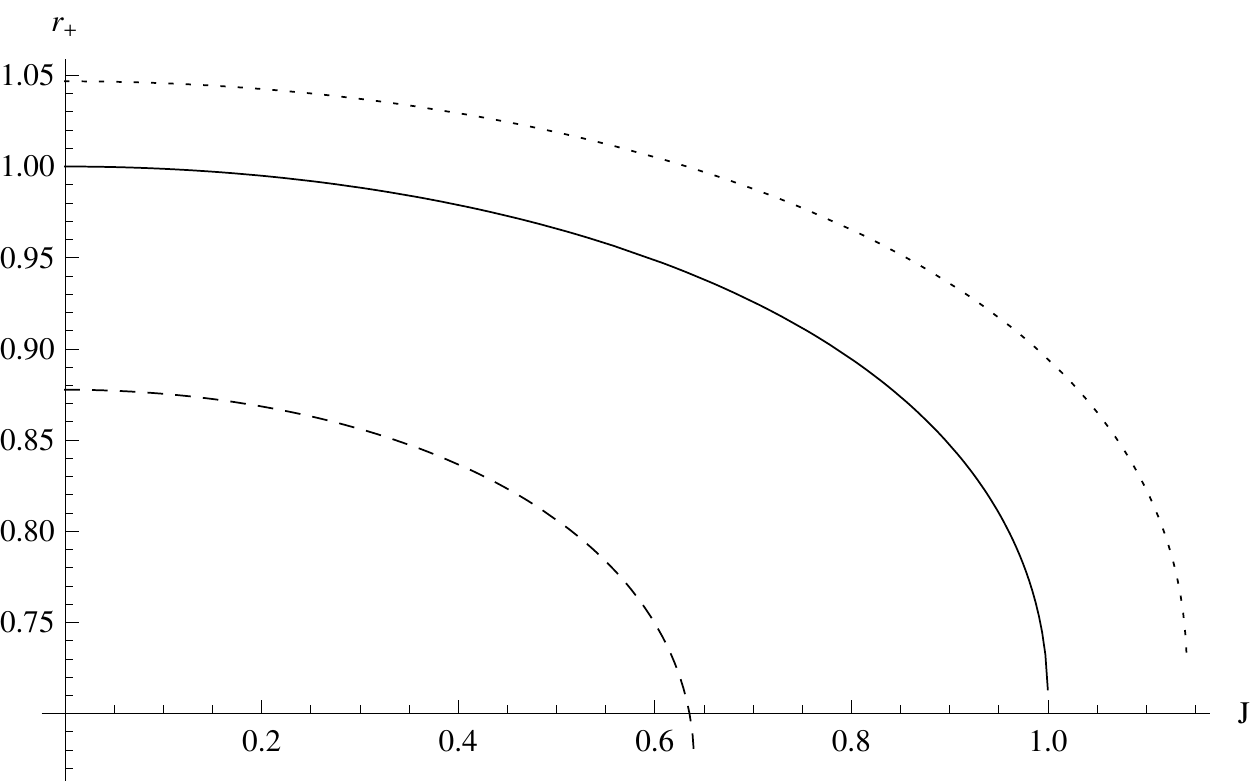}\\
\includegraphics[width=60mm]{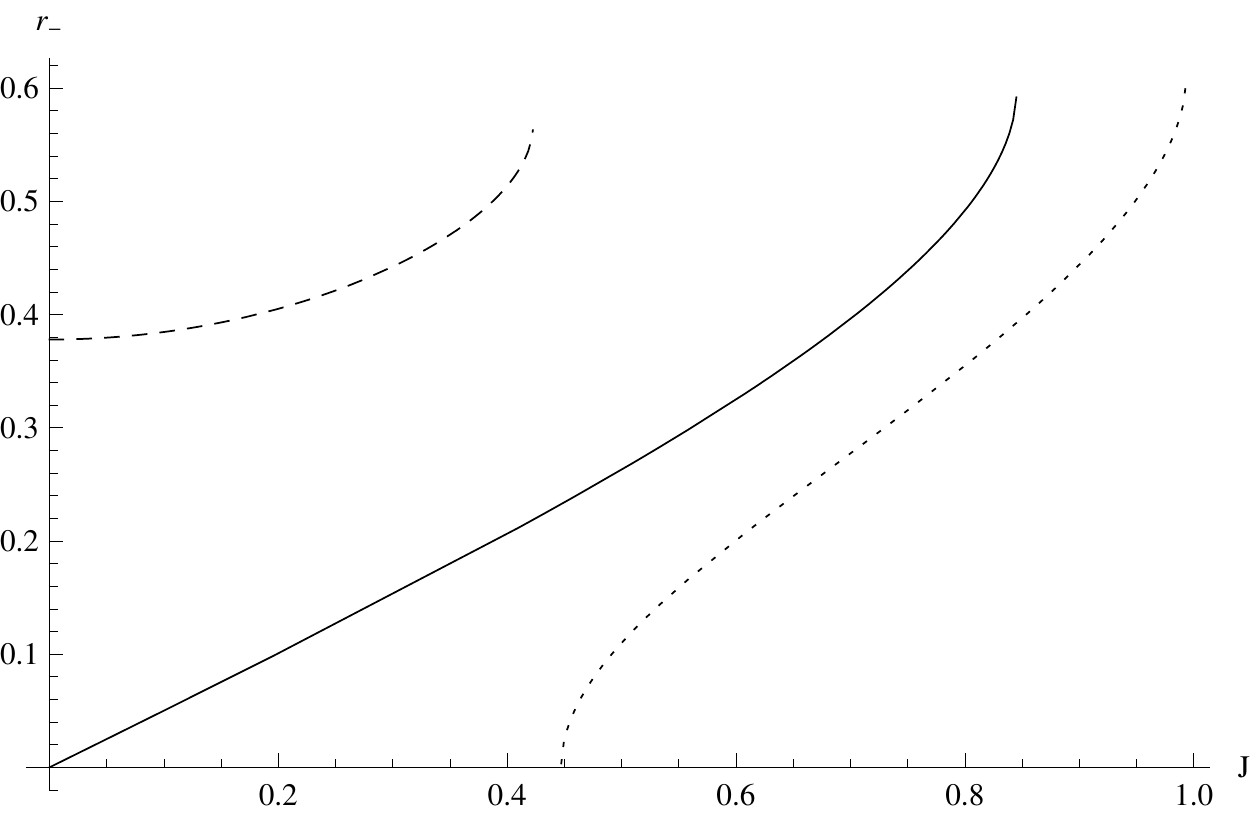}
\includegraphics[width=60mm]{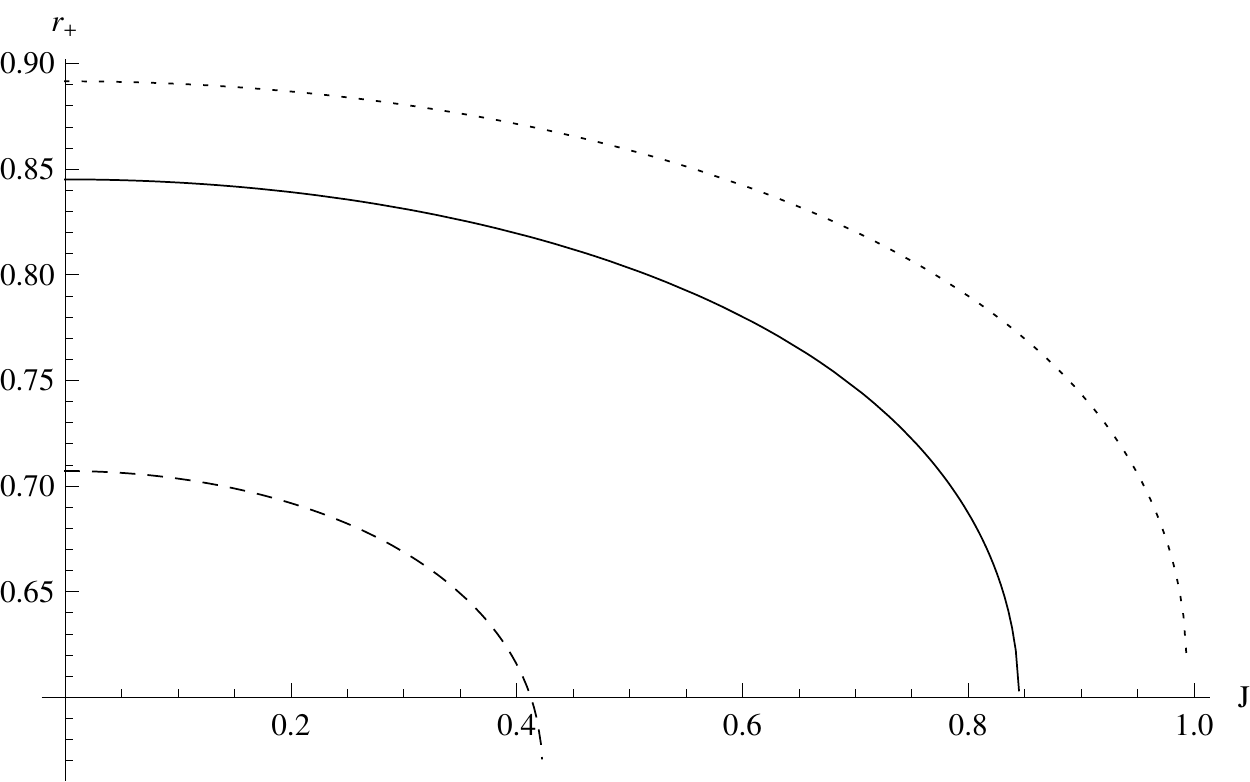}
\end{center}
\caption{The behavior of $r_-$ (left) and $r_+$ (right) as a function of $J$, with $M=1$, $a=1$, $b=1$, and $\Lambda=-1$, dashed lines for $\xi=0.9$, continuous lines for $\xi=1.0$, and dotted lines for $\xi=1.05$. Top panels for $\lambda=0.8$, central panels for $\lambda=1.0$, and bottom panels for $\lambda=1.2$.}
\label{f6}
\end{figure}

\begin{figure}[!h]
\begin{center}
\includegraphics[width=60mm]{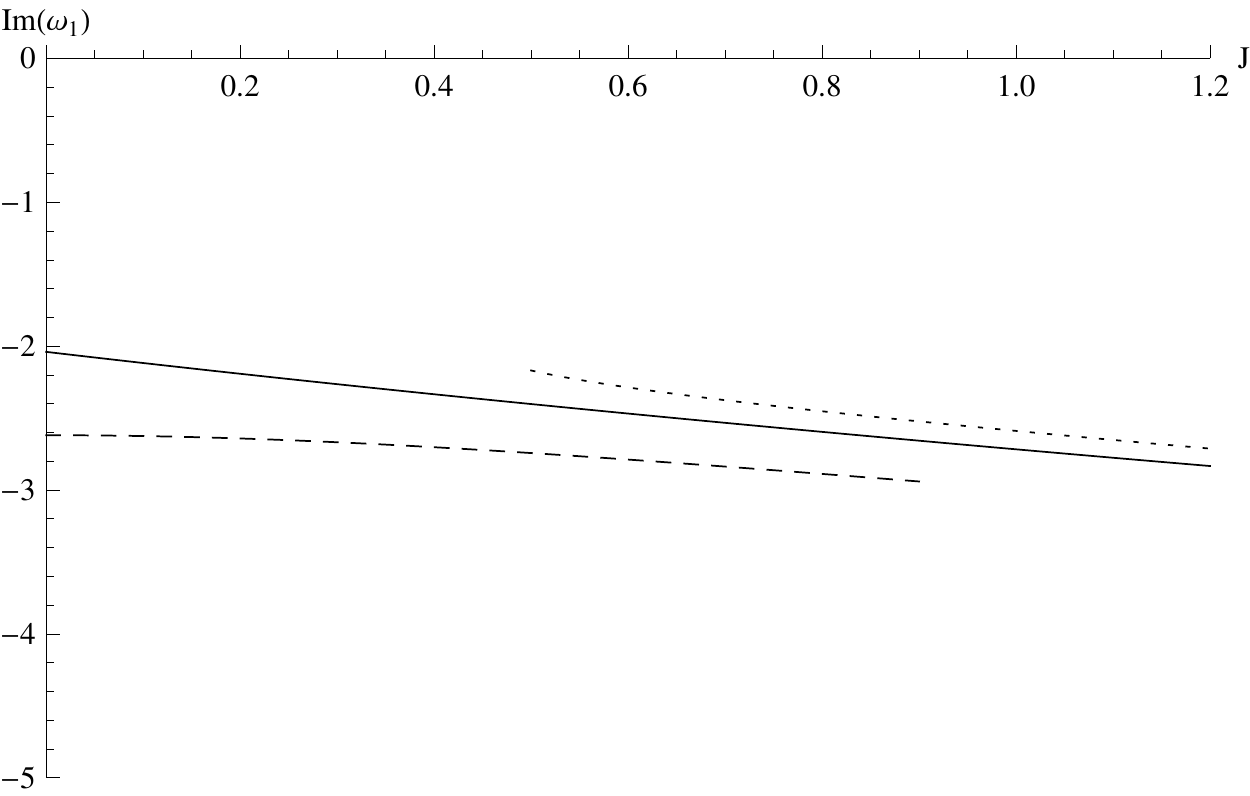}
\includegraphics[width=60mm]{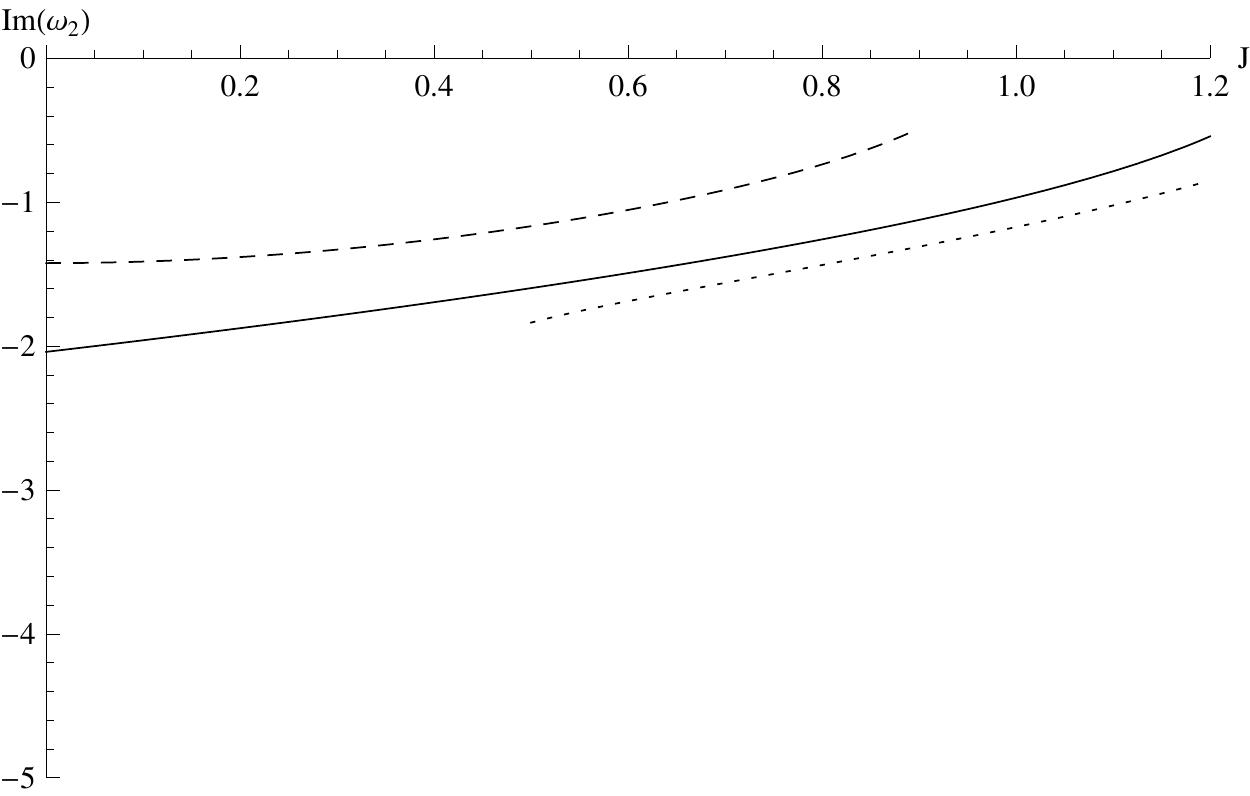}\\
\includegraphics[width=60mm]{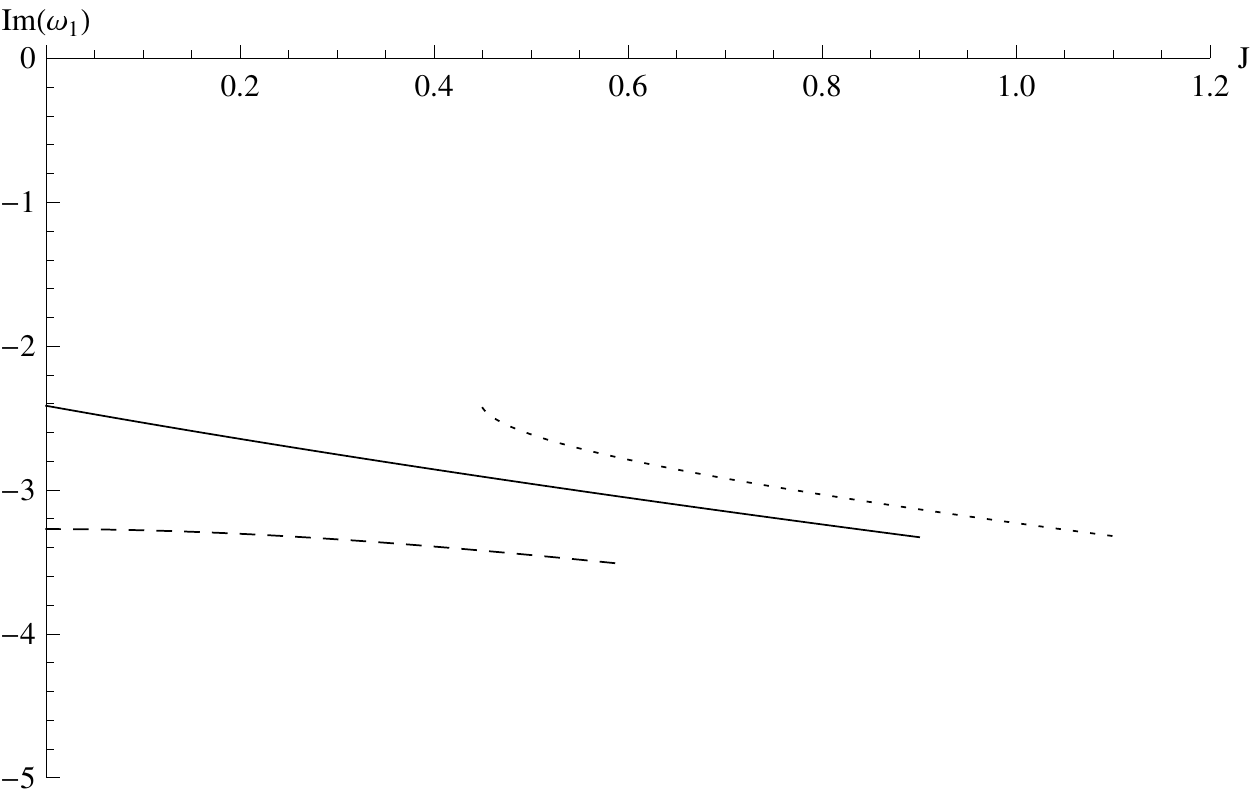}
\includegraphics[width=60mm]{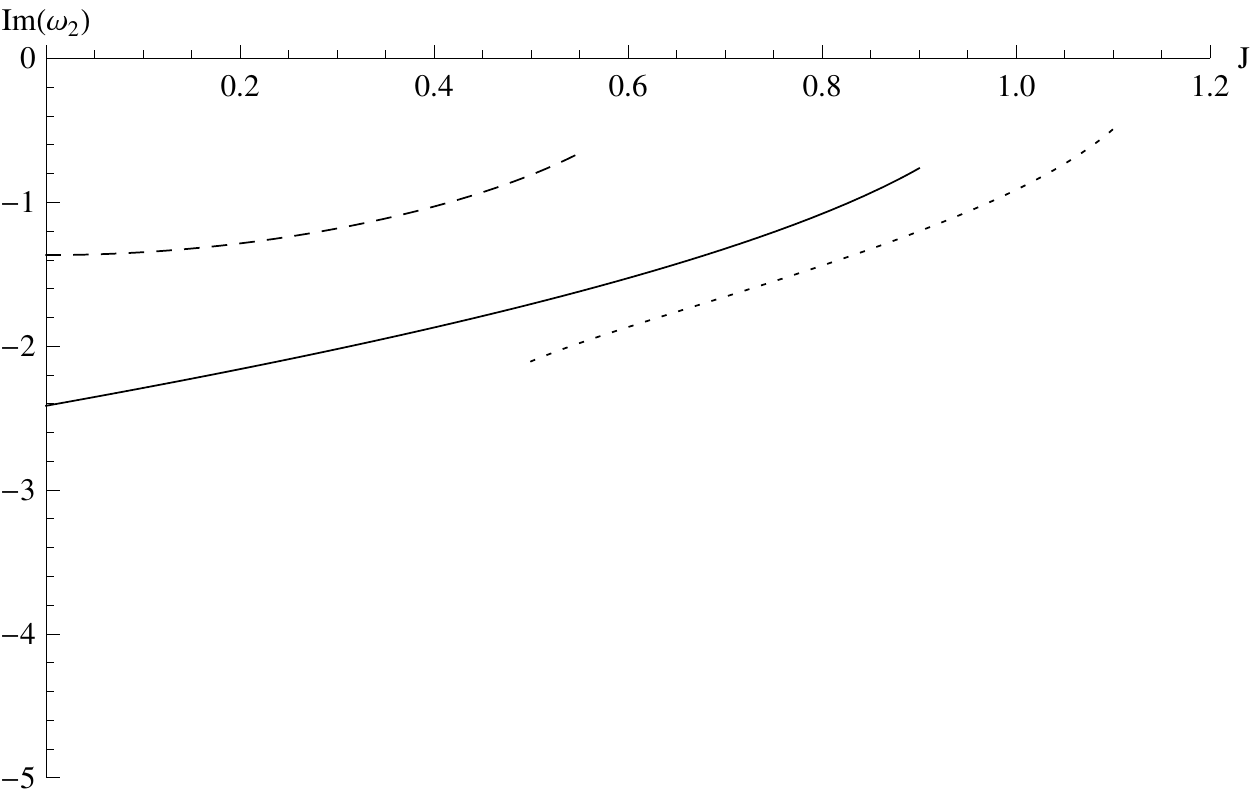}\\
\includegraphics[width=60mm]{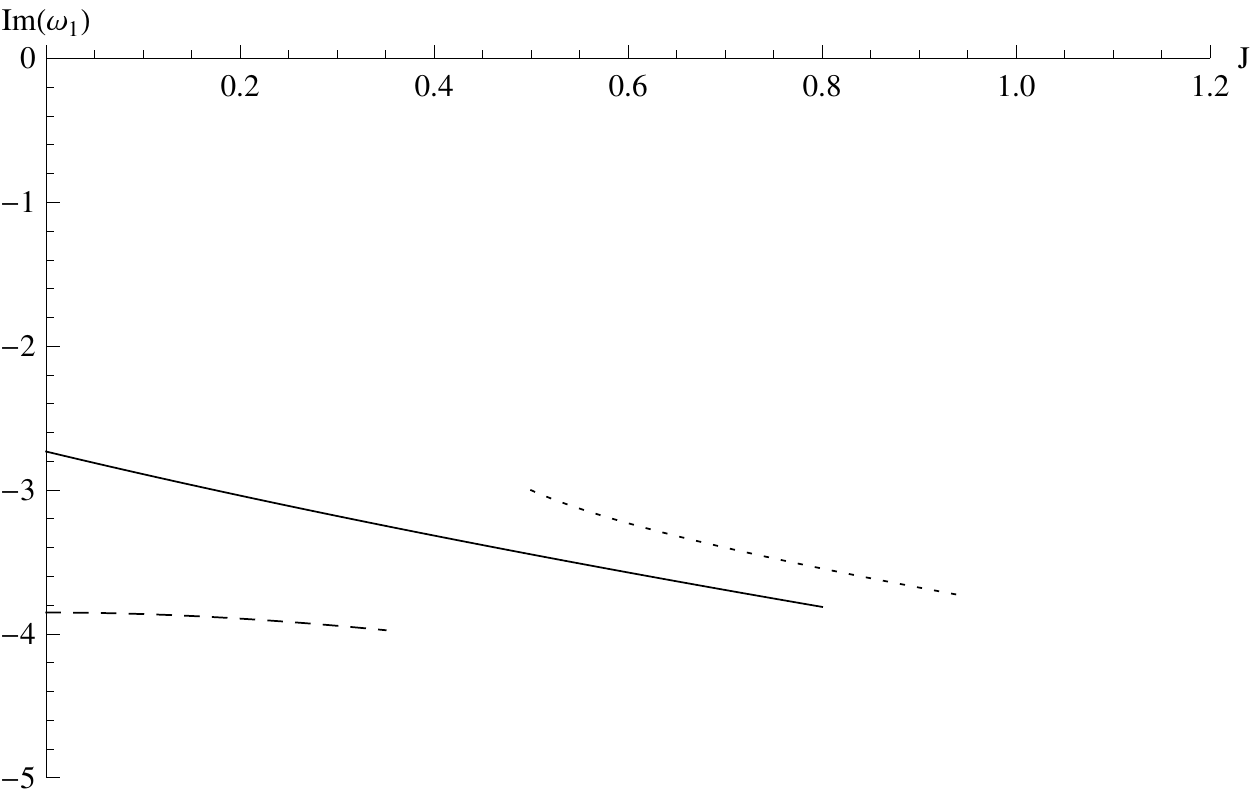}
\includegraphics[width=60mm]{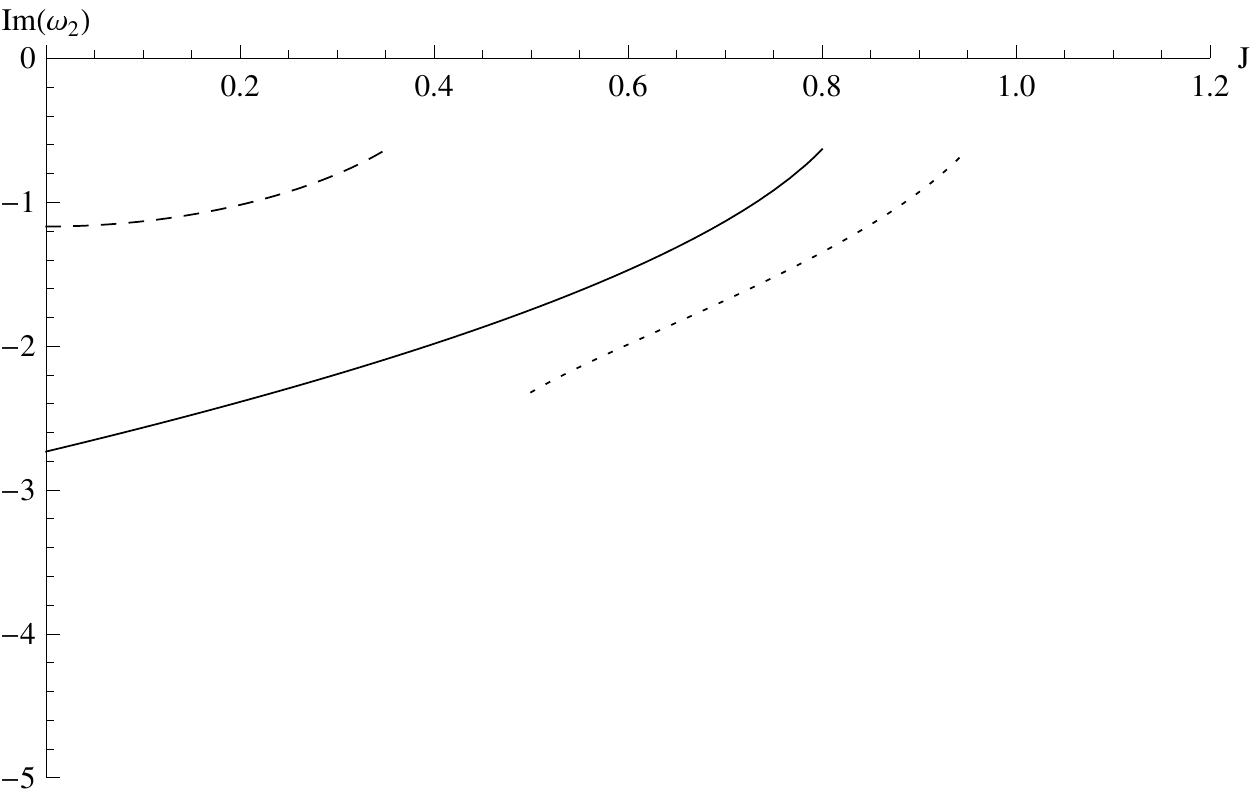}
\end{center}
\caption{The fundamental QNFs  $\omega_1$ (left) and $\omega_2$ (right) as a function of $J$ with $M=1$, $a=1$, $b=1$,  $\Lambda=-1$, $m=1$. Top panels for $\lambda=0.8$, central panels for $\lambda=1$ and bottom panels for $\lambda=1.2$. Dashed lines for $\xi=0.9$, continuous lines for $\xi=1.0$ and dotted lines for $\xi=1.05$.}
\label{f7}
\end{figure}

\begin{figure}[!h]
\begin{center}
\includegraphics[width=90mm]{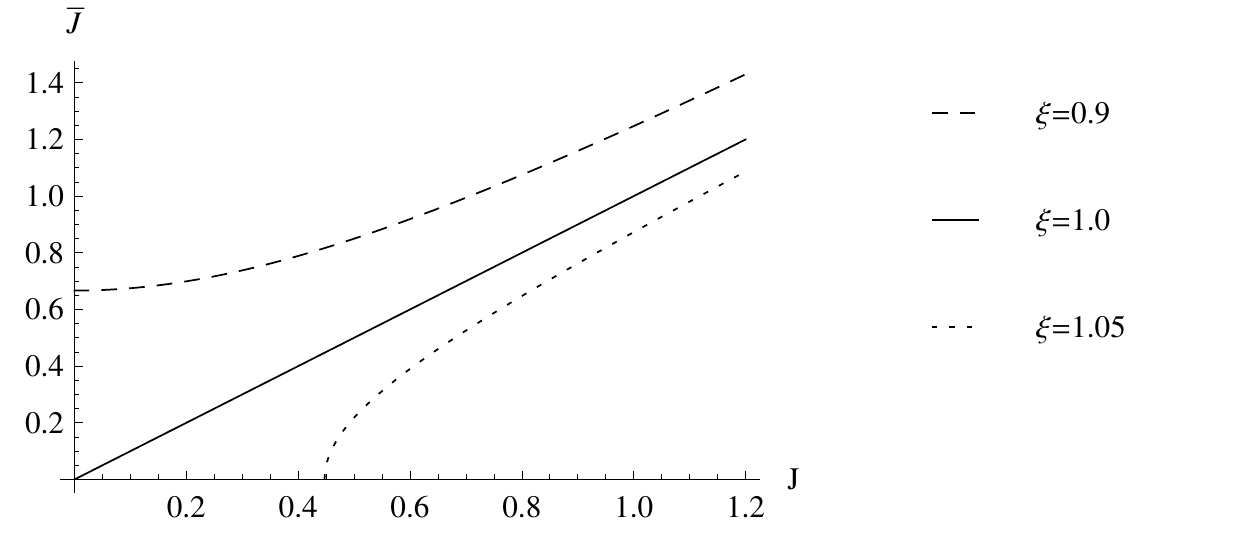}\\
\end{center}
\caption{The behavior of $\bar{J}$ as a function of $J$, with $a=1$. Dashed line for $\xi=0.9$, continuous line for $\xi=1.0$ and dotted line for $\xi=1.05$.}
\label{fJJ}
\end{figure}

Now, in order to observe the behavior of the QNFs (\ref{qnmpp}) and (\ref{qnmp2}), in the range $\xi_e < \xi < \xi_c$, that is, where $r_\pm$ are positive, first we plot $r_\pm$ versus $J$ in Fig. \ref{f6}, for different values of the parameters $\xi$ and $\lambda$,
in order to see for which values of the parameter $J$ the horizons $r_\pm$ are positive. Then, for this range, we plot the imaginary part $Im(\omega)$ of the QNFs in Fig. \ref{f7}, and we observe that for $\omega_1$,
$\abs{Im(\omega_1)}$ increases when the parameter $J$ (or equivalently $\bar{J}$, see Fig. \ref{fJJ}) increases, see left panel of Fig. \ref{f7}, so the relaxation time  decreases. However, for $\omega_2$, the behavior is the opposite. $\abs{Im(\omega_2)}$ decreases when the parameter $J$ increases, see right panel of Fig. \ref{f7}, so the relaxation time  increases. Note that in this range $Re(\omega)$ is null. Also, the sectors $T_R$ and $T_L$ of the conformal field theory are well defined. Furthermore, $\abs{Im(\omega_1)}$ decreases and $\abs{Im(\omega_2)}$  increases, when the coupling constant $\xi$ increases; so, the relaxation time increases and decreases, respectively.


\newpage

\begin{figure}[!h]
\begin{center}
\includegraphics[width=60mm]{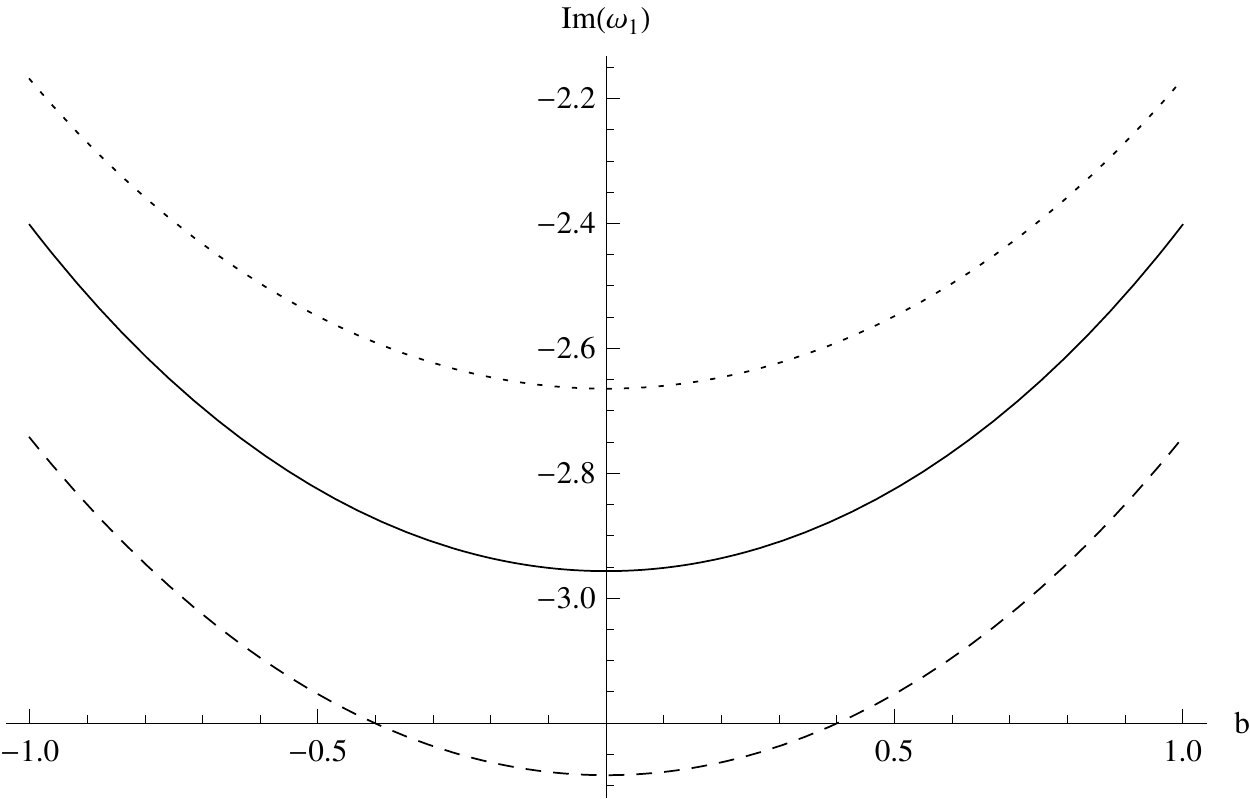}
\includegraphics[width=60mm]{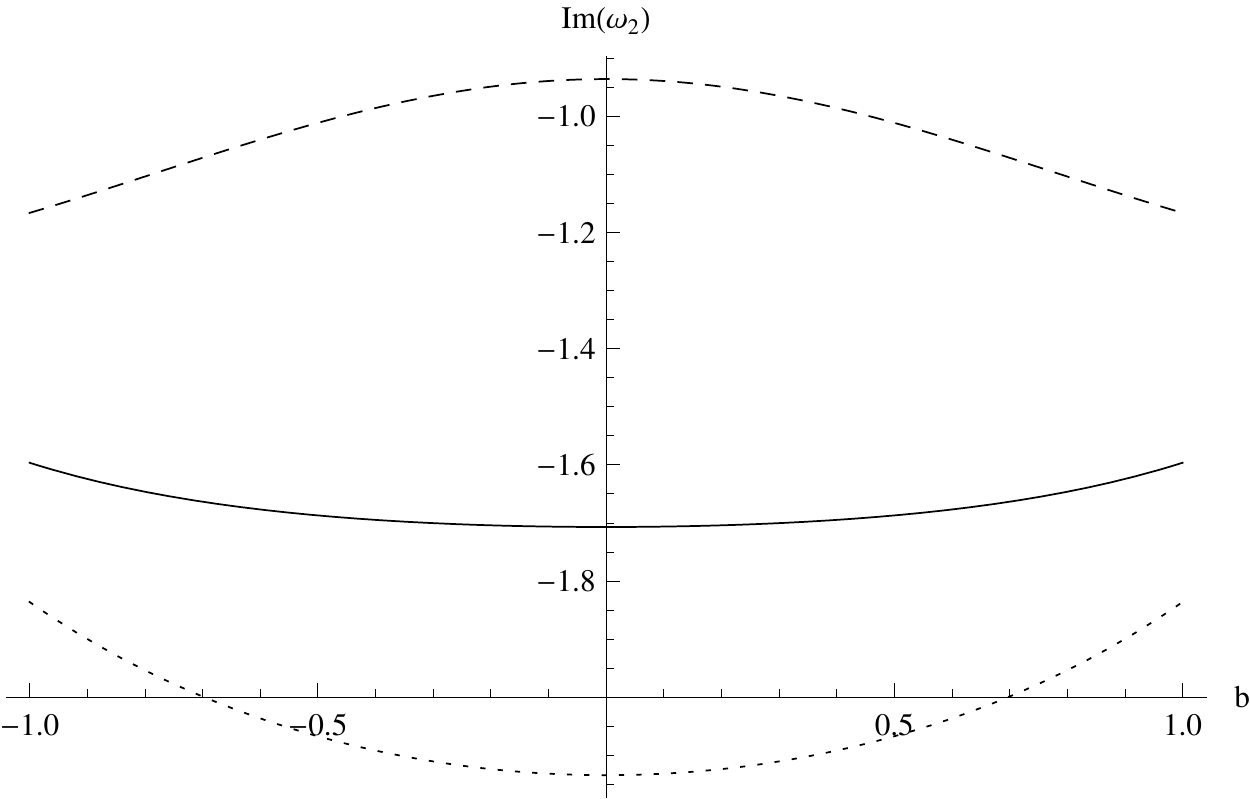}\\

\includegraphics[width=60mm]{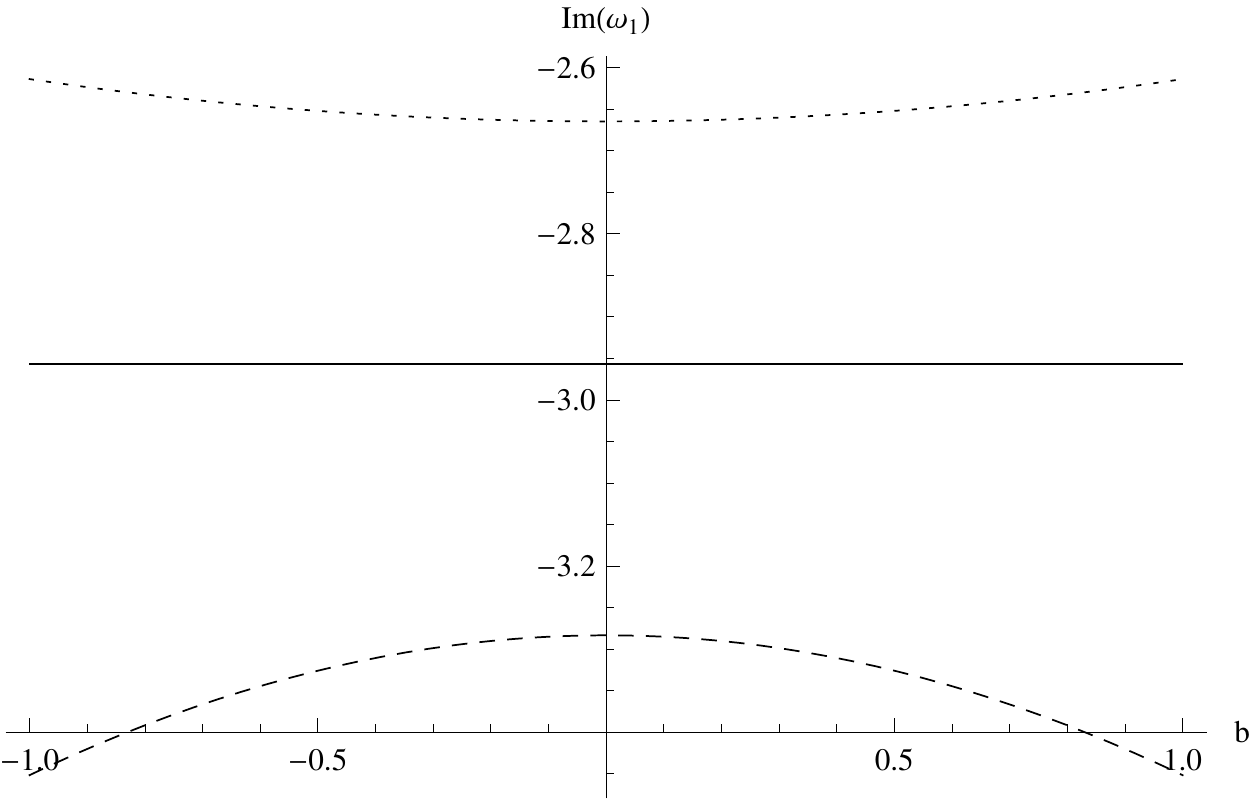}
\includegraphics[width=60mm]{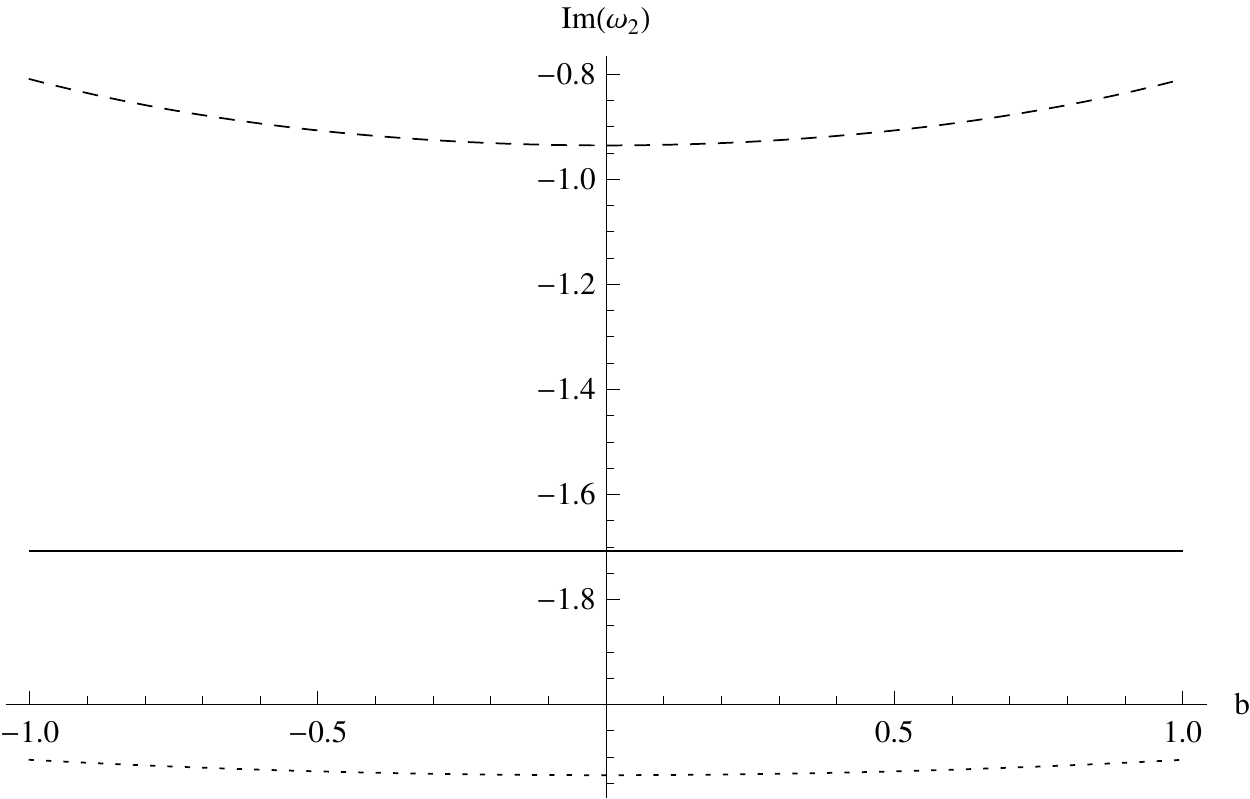}\\

\includegraphics[width=60mm]{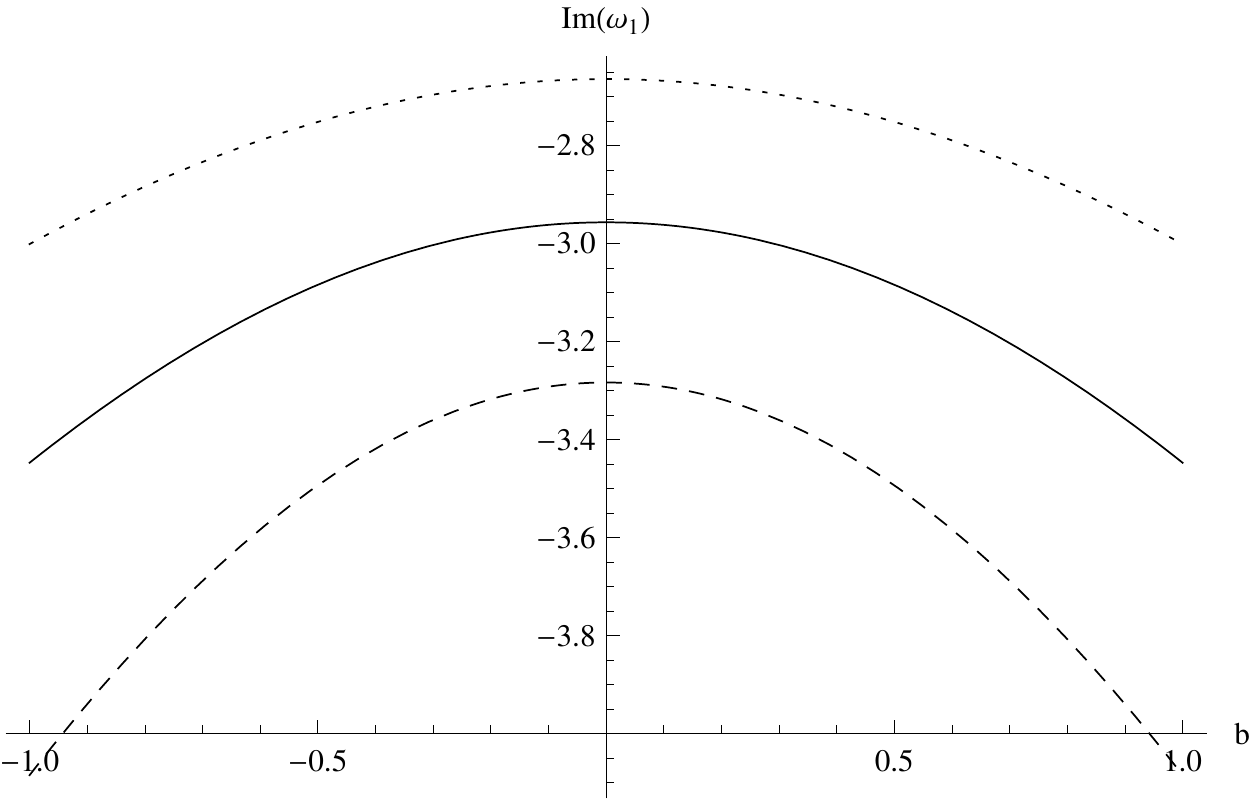}
\includegraphics[width=60mm]{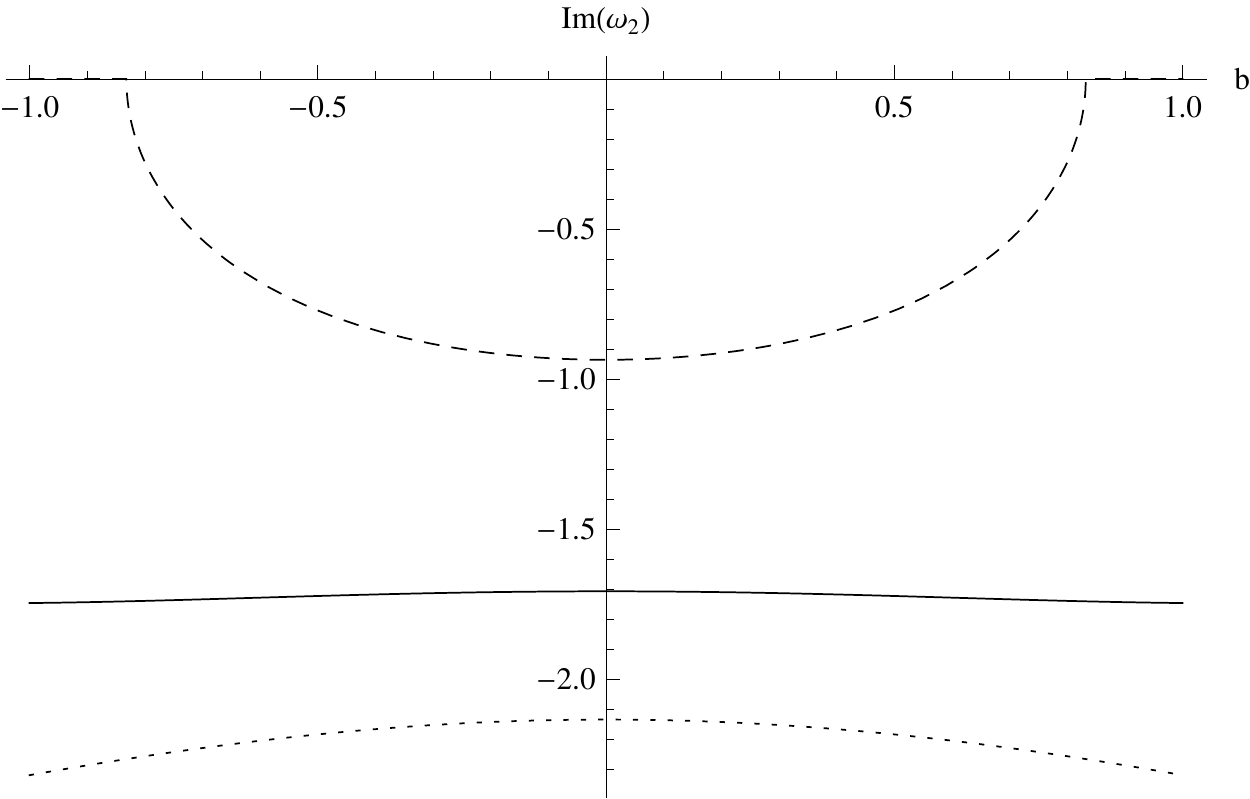}
\end{center}
\caption{The fundamental QNFs $\omega_1$ (left) and $\omega_2$ (right) as a function of $b$ with $M=1$, $a=1$,  $\Lambda=-1$, $J=0.5$, and $m=1$. Top panels for $\lambda=0.8$, central panels for $\lambda=1$, and bottom panels for $\lambda=1.2$. Dashed lines for $\xi=0.9$, continuous lines for $\xi=1.0$, and dotted lines for $\xi=1.05$.}
\label{f8}
\end{figure}

In Fig. \ref{f8} we plot the fundamental QNFs as a function of $b$. We observe that for the BTZ black hole, the central panel with $\xi=1$, the imaginary part of the fundamental QNF is constant; however, for a asymptotical misalignment of the aether with the timelike Killing vector, $b \neq 0$, the fundamental QNFs depend on $b$.  $\abs{Im(\omega_1)}$ increases when $\xi$ decreases. For $\abs{Im(\omega_2)}$ the behavior is the opposite since it decreases when $\xi$ decreases.

\subsubsection{$\xi > \xi_c$}

\begin{figure}[!h]
\begin{center}
\includegraphics[width=60mm]{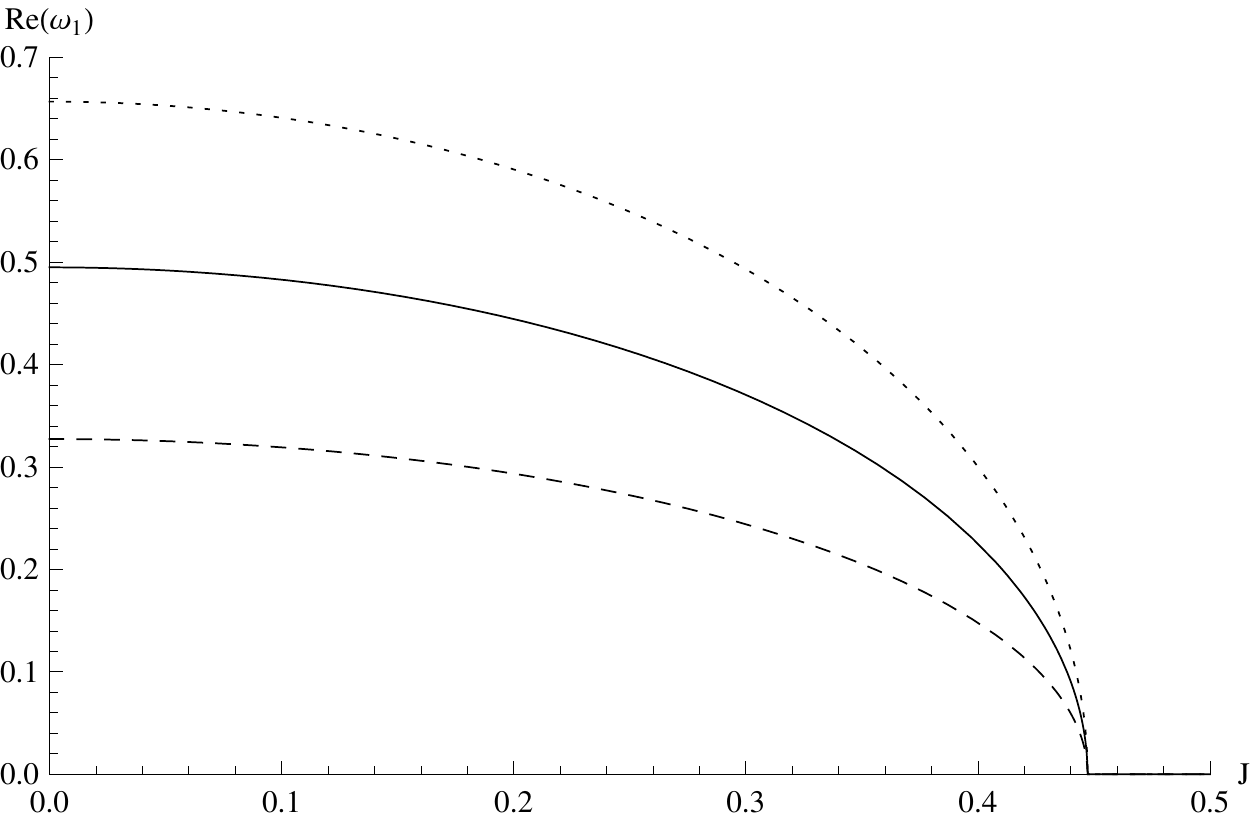}
\includegraphics[width=60mm]{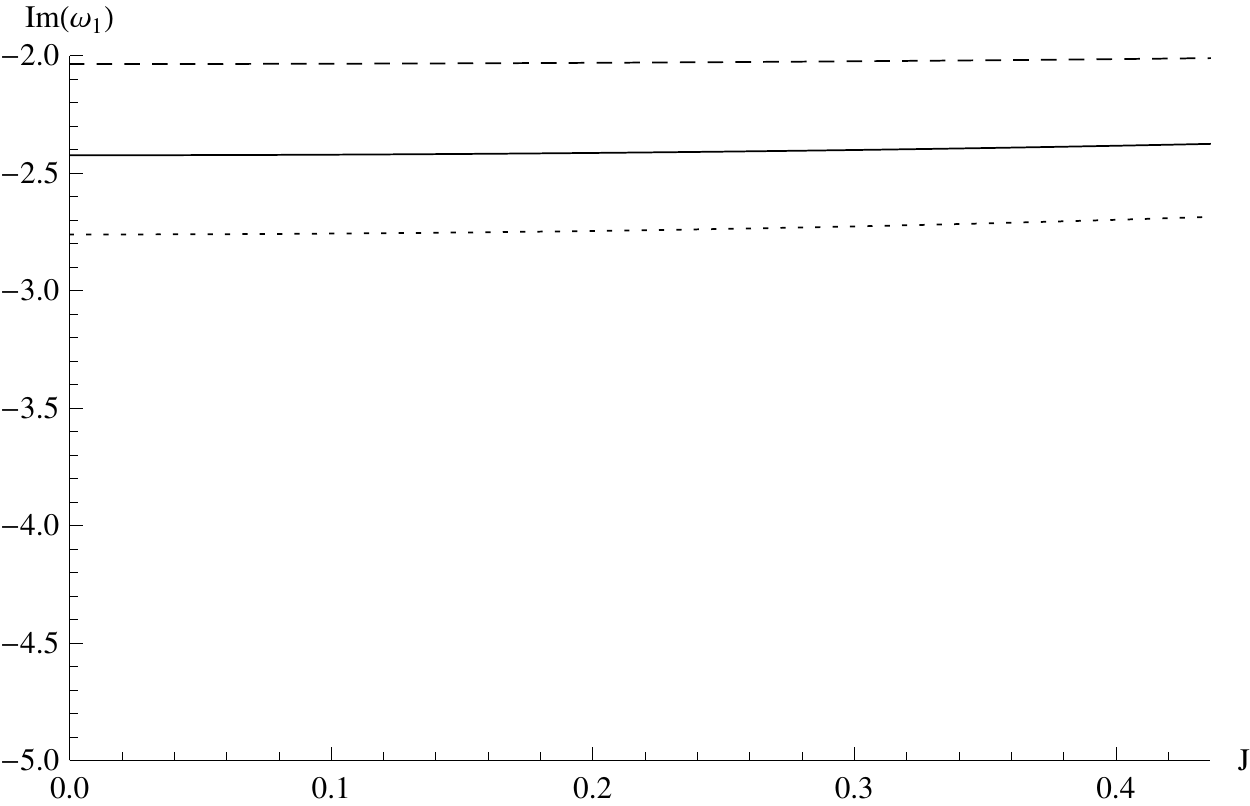}
\includegraphics[width=60mm]{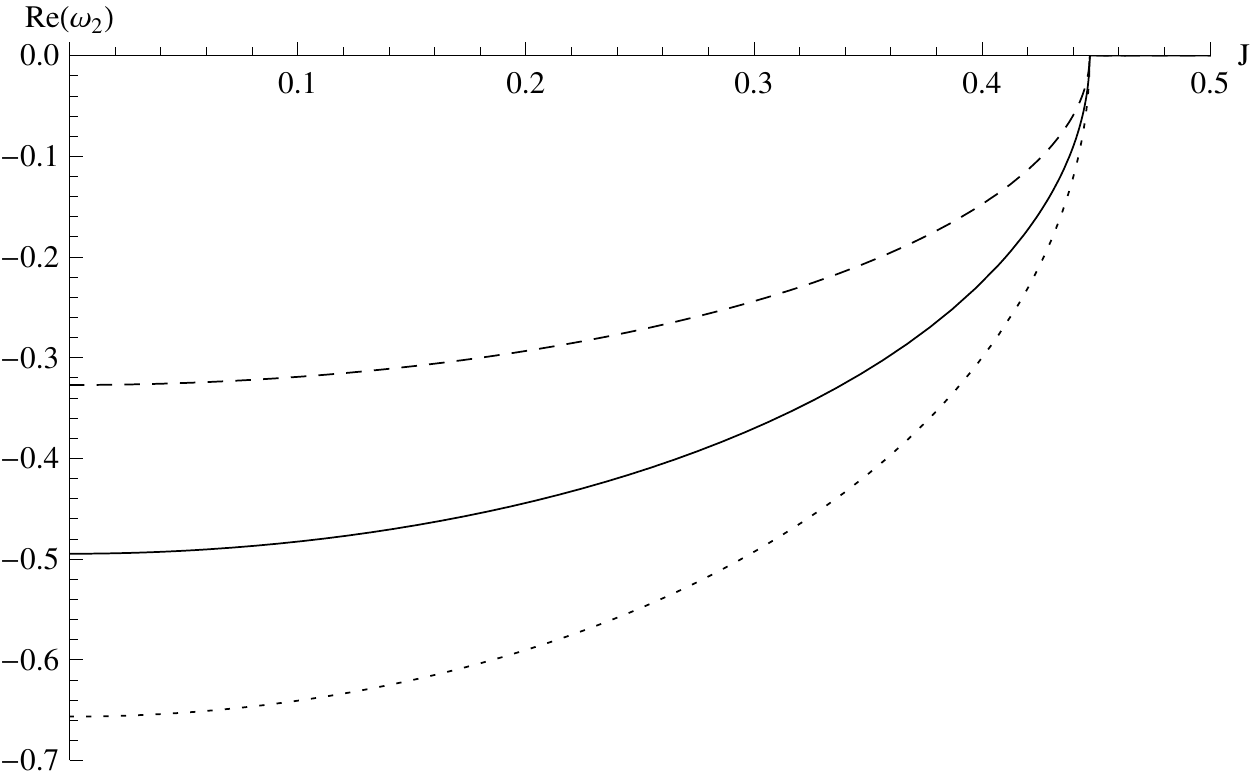}
\includegraphics[width=60mm]{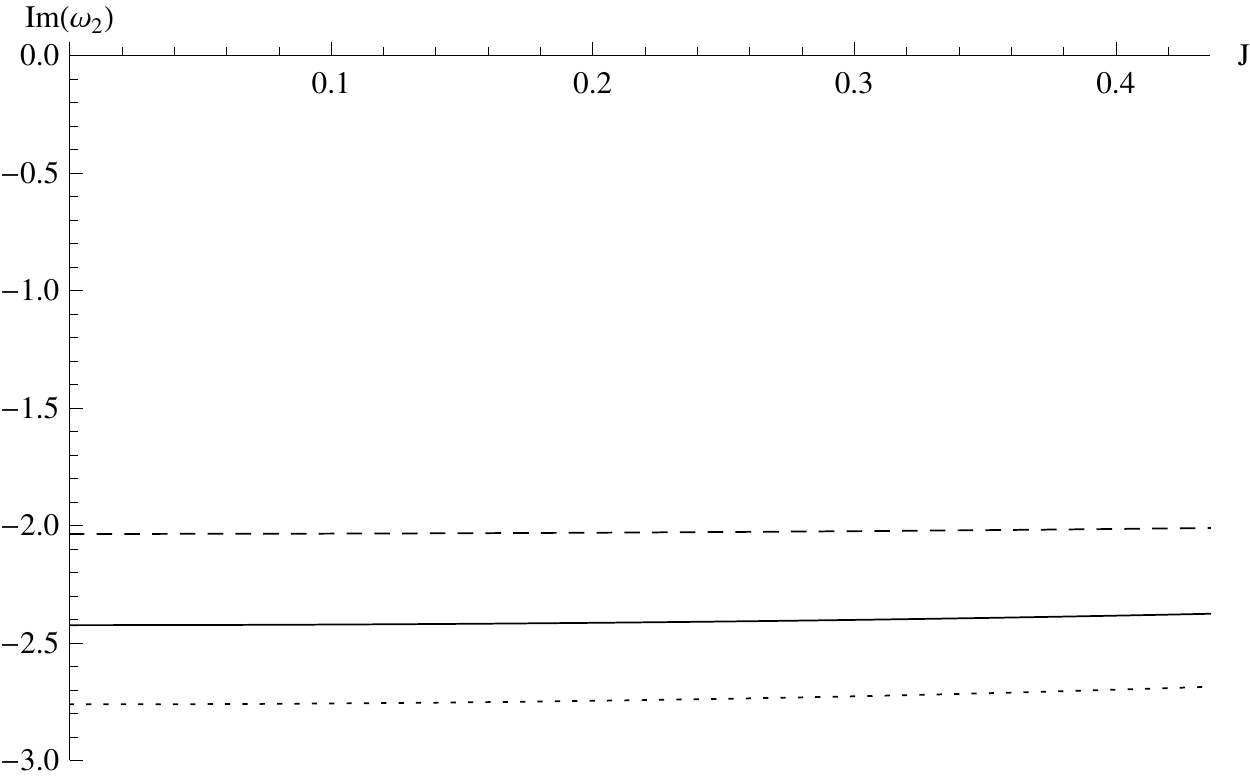}
\end{center}
\caption{The real (left) and imaginary (right) part of the fundamental quasinormal frequency ($\omega_1$) as a function of $J$, for $0<J<\sqrt{0.2}$, with $\xi=1.05$, $M=1$, $a=1$, $b=1$, $m=1$. Dashed line for $\lambda=0.8$, continue line for $\lambda=1$, and dotted line for $\lambda=1.2$.}
\label{friw1}
\end{figure}

As mentioned, for $\xi > \xi_c$ there is only one horizon. $r_+>0$ and $r_-$ become imaginary. This occurs for $\xi > (J^2+4a^2)/(4a^2)$, and consequently there is a gap in $J$, see Fig. \ref{f6}, left panel, and for which $\bar{J}^2$ is negative, see Fig. \ref{fJJ}, which occurs when $J^2<4a^2(\xi-1)$. In Fig. \ref{friw1}, we plot the fundamental QNFs for the range of values of $J$ in which it is positive, and there is only one horizon. So, we observe that the fundamental QNFs acquire a real part, with $Re(\omega_1)=-Re(\omega_2)$,  $\abs{Re(\omega)}$ decreases when $J$ increases,  and $Im(\omega_1)=Im(\omega_2)$, and it is negative. In this case, the two branches have converge to one branch and when the coupling constant $\xi$ increases  $\abs{Im(\omega)}$ decreases, see Fig. \ref{omegaxi}.

\subsubsection{$\xi=\xi_c$}

\begin{figure}[!h]
\begin{center}
\includegraphics[width=60mm]{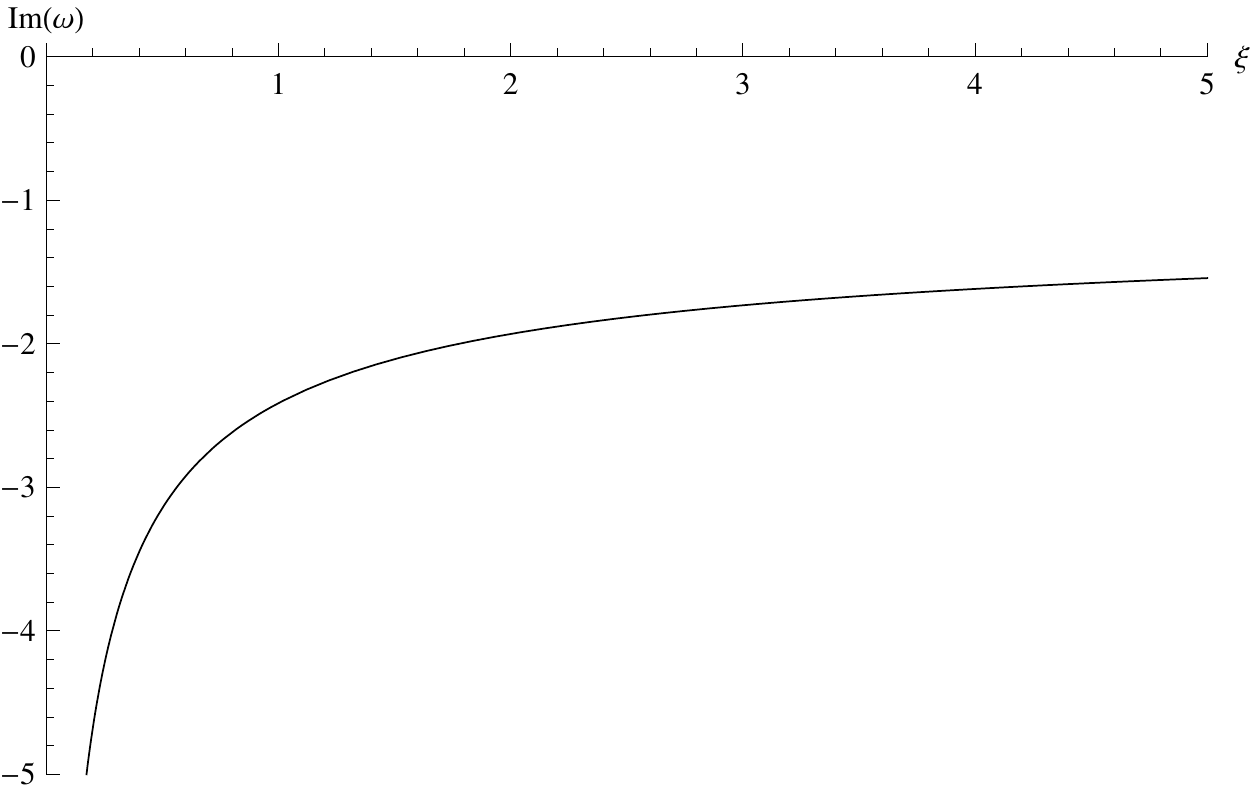}
\end{center}
\caption{The fundamental QNFs $Im(\omega)$ as a function of $\xi_c$ with $M=1$, $b=1$,  $\Lambda=-1$, $\lambda=1$, $m=1$, and $J=0.4$.}
\label{fomega}
\end{figure}

Finally, for $\xi=\xi_c$, that is  $r_-=0$, the two sectors converge, this occurs for $\bar{J}=0$, that is  $J^2+4a^2(1-\xi)=0$. In this case the QNFs are given by
\begin{equation}\label{qnmpp0}
\omega_1=\omega_2=-i \abs{\bar{\Lambda}}r_+ \left(\sqrt{1-4C}+2n+1\right)\,,
\end{equation}\\
where $r_{+}^2=-\frac{M}{\bar{\Lambda}}$.
In Fig. \ref{fomega} we show the behavior of the fundamental QNFs as a function of $\xi=\xi_c$. We observe that $\abs{Im(\omega)}$   decreases when $\xi_c$ increases whereas $Re(\omega)$ is null. So, according to the gauge/gravity duality, the relaxation time in order to reach the thermal equilibrium increases.

\newpage

\begin{figure}[!h]
\begin{center}
\includegraphics[width=55mm]{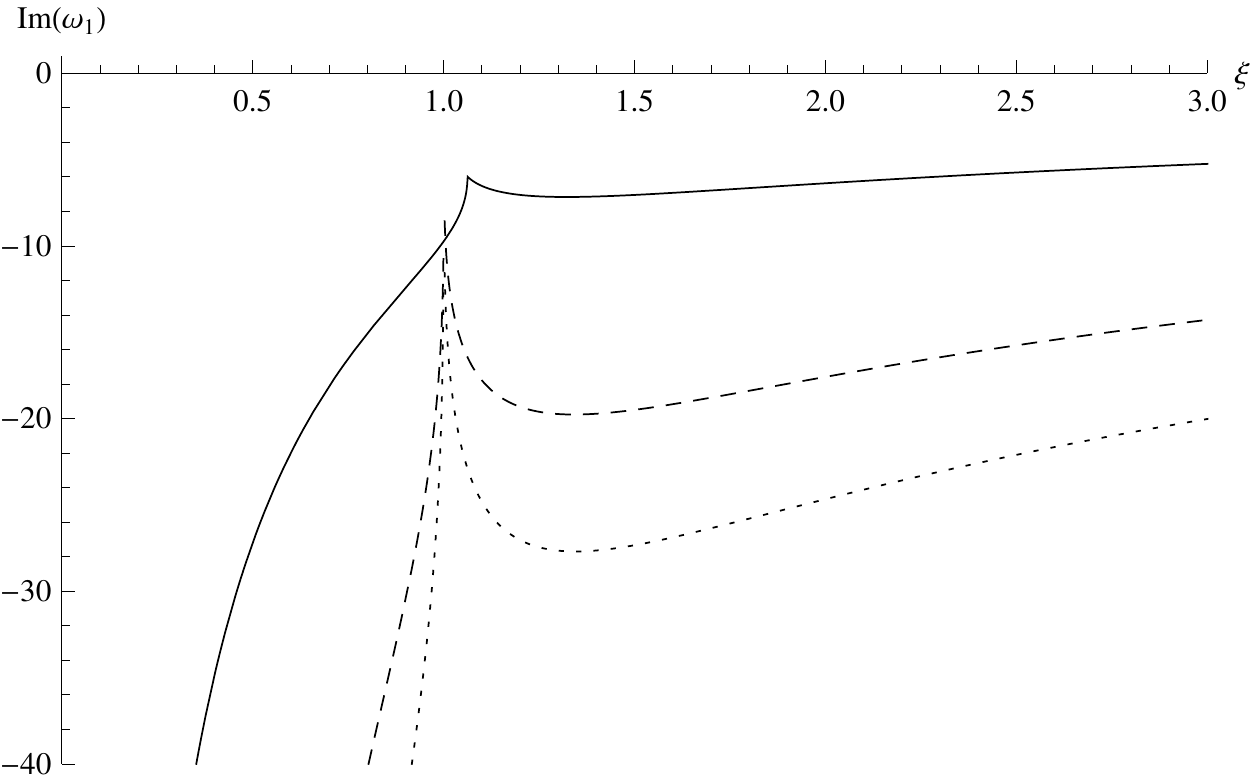}
\includegraphics[width=82mm]{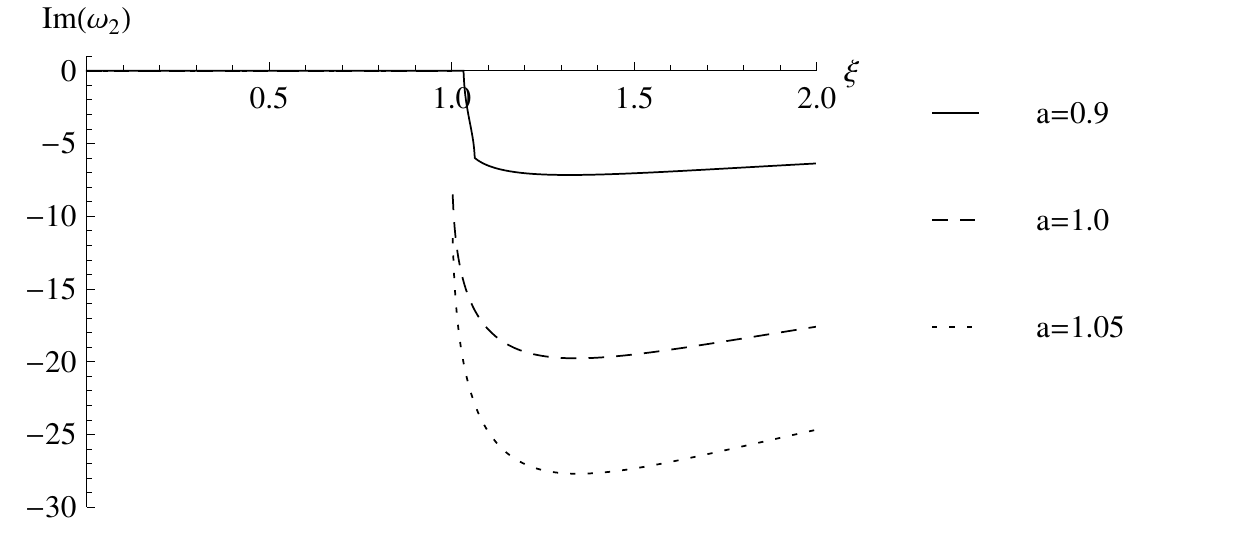}
\includegraphics[width=55mm]{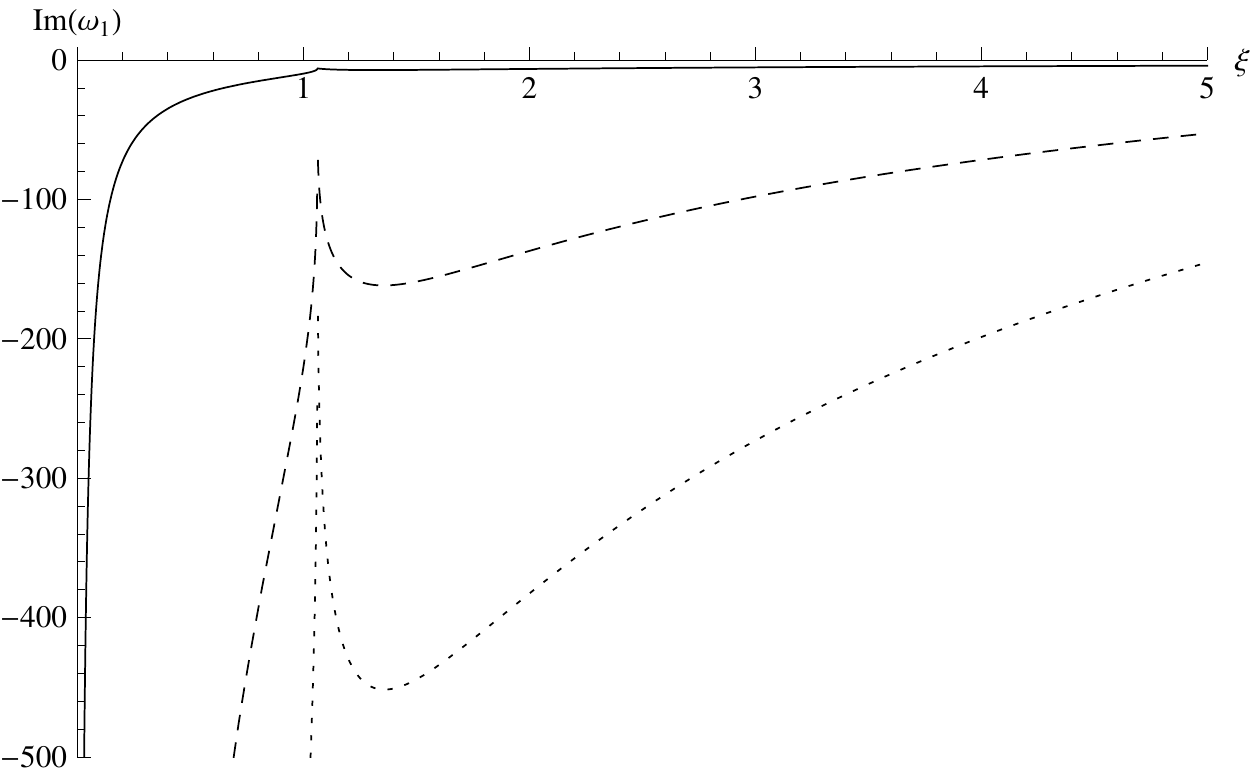}
\includegraphics[width=82mm]{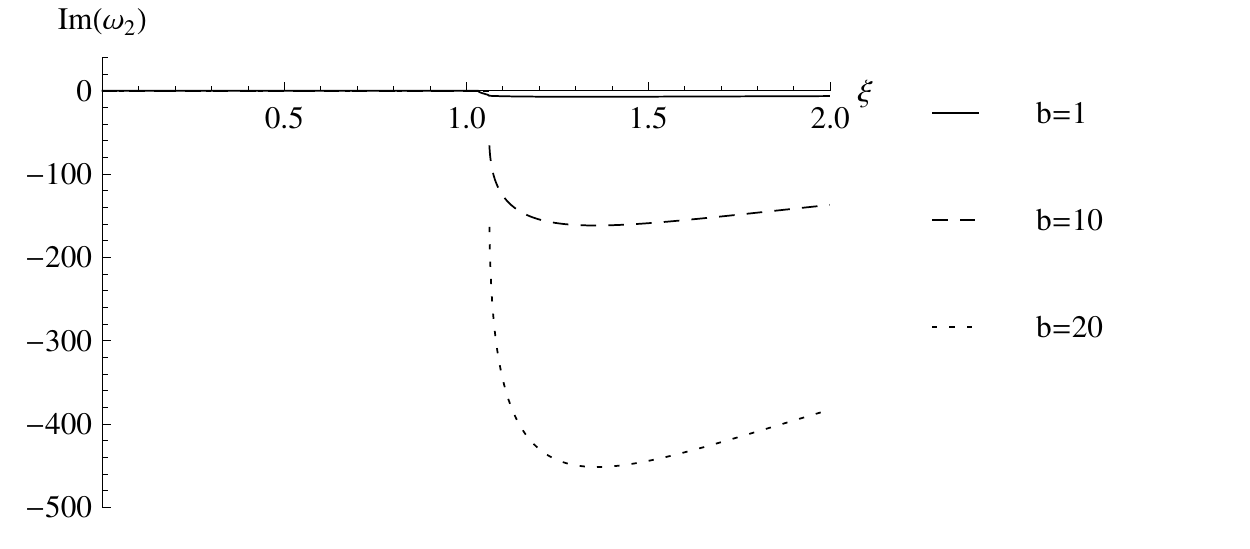}
\end{center}
\caption{The imaginary  part of the fundamental QNFs $\omega_1$ (left panels) and $\omega_2$ (right panels) as a function of $\xi$, with $J=0.5$, $M=1$, $m=1$, and $\lambda=5$, for different values of the constant $a$ (top panels, with $b=1$) and $b$ (bottom panels, with $a=1$).}
\label{omegaab}
\end{figure}

Finally, we plot in Fig. \ref{omegaab} $Im(\omega)$ for different values of the constants $a$ and $b$. We observe that for the range $\xi>\xi_e$, $\abs{Im(\omega)}$ increases when the constants $a$ or $b$ increase, so the relaxation time decreases.

\emph{Neumann boundary conditions.} The frequencies found above for the scalar perturbation have been obtained by imposing the vanishing Dirichlet boundary condition at infinity.  It is known that the Dirichlet boundary condition does not lead to any quasinormal modes for $m^2 < 0$.
However, it is also possible to find a second set of QNFs, for negative mass squared, if we consider that the flux of the scalar field vanishes at infinity or vanishing Neumann boundary condition at infinity,  which allows us to describe tachyons. Furthermore, it was shown that for negative mass squared there are two sets of dual operators  $\Delta_+$ and $\Delta_-$, where the second set of QNFs  matches exactly the dual operators with $\Delta_-$ \cite{Birmingham:2001pj}.
So, by using the condition that the flux, which is given by
\begin{equation}
F=\frac{\sqrt{-g}g^{rr}}{2i}(\psi^{*}\partial_r \psi-\psi\partial_r \psi^*)\,,
\end{equation}
vanishes at asymptotic infinity, we obtain for $\beta=\beta_-$ and $0>m^2>\bar{\Lambda}$, that the flux vanishes if  $(a)|_{\beta_-}=-n$ or $(b)|_{\beta_+}=-n$, which leads to
\begin{equation}\label{qnmp2f}
\omega=-i \abs{\bar{\Lambda}} \left(-\sqrt{1-4C}+2n+1\right) (r_+ - r_-)\,,
\end{equation}
\begin{equation}\label{qnmppf}
\omega=-i \abs{\bar{\Lambda}} \left(-\sqrt{1-4C}+2n+1\right) (r_- + r_+)\,,
\end{equation}
respectively.

\subsection{Massive scalar field}

\begin{figure}[!h]
\begin{center}
\includegraphics[width=80mm]{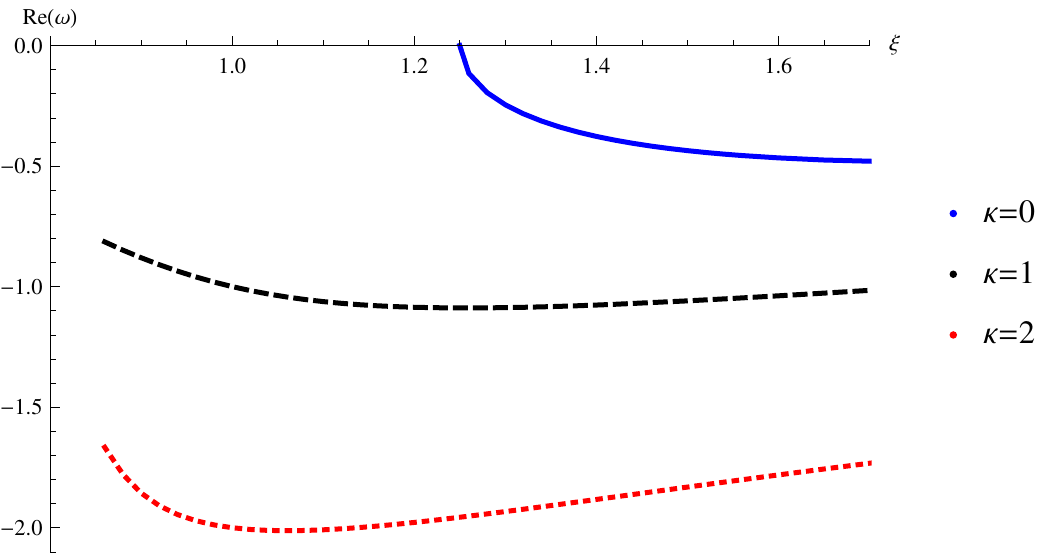}
\includegraphics[width=80mm]{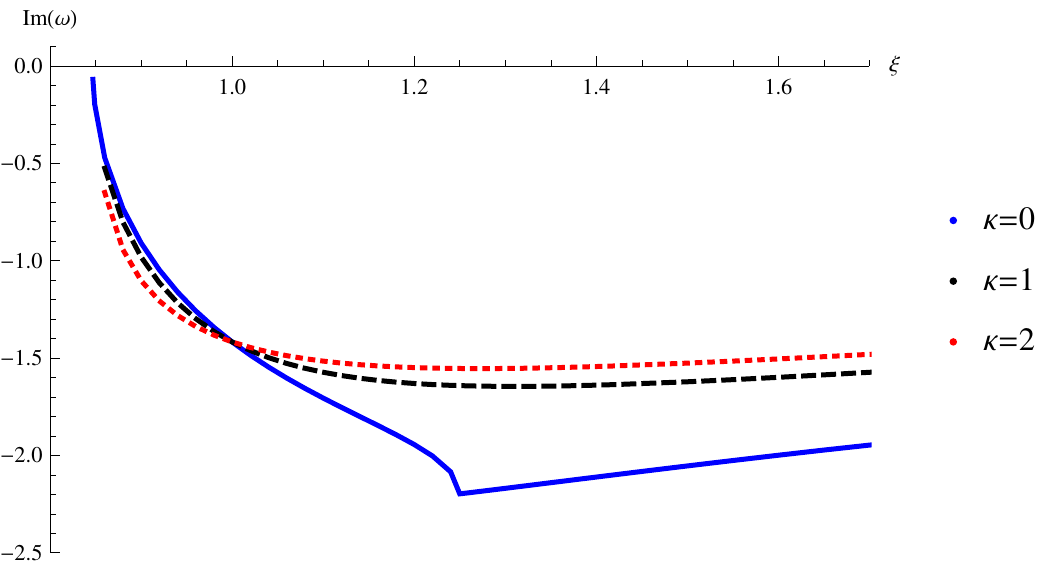}
\end{center}
\caption{The fundamental QNFs of the branch $\omega_2$ for $a=1$, $b=1$, $\lambda=1$, $\Lambda=-1$, $J=1$, $m=0.1$ and $M=1.5$ as a function of $\xi$ for $\kappa= 0,1,2$. Left panel for $Re(\omega)$, and right panel for $Im(\omega)$.}
\label{sp}
\end{figure}

For massive scalar field, the Klein-Gordon equation
can be rewritten as
\begin{equation}\label{radial12222}\
z(1-z)\partial_{z}^2R(v)+\left(1-z\right)\partial_{z}R(z)+
\left[A+\frac{B}{z}+\frac{C}{1-z}+\frac{D}{\bar{a}-z}\right]R(z)=0 \,,
\end{equation}
where
\begin{equation}
\label{A}
A=
-\frac{(\omega r_--\kappa r_+)^2}{4\bar{\Lambda}^2(r_+^2-r_-^2)^2}~,~
B=
\frac{(\omega r_+-\kappa r_-)^2}{4\bar{\Lambda}^2(r_+^2-r_-^2)^2}~,~
C=\frac{m^2}{4\bar{\Lambda}} ~,~
D=\frac{\kappa ^2 \left((-1+\bar{a})^2 J^2+4 \bar{a} \bar{\Lambda} (r_-^2-r_+^2)^2\right)}{16 \bar{a} \bar{\Lambda}^2 r_-^2 (r_-^2-r_+^2)^2}~,~
\end{equation}
and
\begin{equation}
\bar{a}=r_+^2/r_-^2 ~.
\end{equation}
Under the decomposition $R(z)=z^\alpha(1-z)^\beta(\bar{a}-z)K(z)$,
with
\begin{equation}
\alpha=\pm i \sqrt{B} ~,\,
\beta=\frac{1}{2} \left(1\pm\sqrt{1-4 C}\right) ~,
\end{equation}
Eq.(\ref{radial12222}) can be written as
\begin{equation}\label{radial1222}\
\partial_{z}^2K(z)+\left(\frac{1+2\alpha}{z}-\frac{2\beta}{1-z}-\frac{2}{\bar{a}-z}\right)\partial_{z}R(z)+
\frac{1}{(z)(1-z)(\bar{a}-z)}\left(q +\epsilon_1 \epsilon_2 z\right)R(z)=0 \,,
\end{equation}
where
\begin{equation}
   q = -1-B+C+D-2 \alpha -\alpha ^2-\beta +\beta ^2+\bar{a} \left(A-(\alpha +\beta )^2\right)~,
\end{equation}
\begin{equation}
   \epsilon_1 = -\sqrt{A}+(1+\alpha +\beta )~,\,
   \epsilon_2 = \sqrt{A}+(1+\alpha +\beta )~,
\end{equation}
that corresponds to Heun's differential equation. The condition $2=\epsilon_1+\epsilon_2-(1+2\alpha)-2\beta+1$, ensures regularity of the point at $\infty$, and $q$ corresponds to the accessory parameter. Heun's equation has four regular singular points: $0,1,a$, and $\infty$ with exponents $(0,-2\alpha)$, $(0, 1-2\beta)$, $(0,-1)$, and $(\epsilon_1,\epsilon_2)$. Now, in order to obtain the QNFs, we proceed to perform a numerical analysis by using the pseudospectral Chebyshev method \cite{Boyd}, which has been applied to compute the QNFs in other geometries, for instance see \cite{Gonzalez:2017shu, Gonzalez:2018xrq, Finazzo:2016psx}. In Fig. \ref{sp} we plot the numerical results obtained for the real and imaginary parts of the fundamental QNF of the branch $\omega_2$ as a function of $\xi$ for different values of $\kappa$. We observe that for $\xi>1$, the absolute value of the imaginary part decreases as $\kappa$ increases; however, for $\xi <1$, the behavior is the opposite; the absolute value of the imaginary part increases as $\kappa$ increases. Note that, for the cases analyzed, the QNFs have a negative imaginary part, which ensures that the propagation of scalar fields is stable in this background.

\newpage

\subsection{Case: $(-1+\bar{a})^2 J^2+4 \bar{a} \bar{\Lambda} \left(r_-^2- r_+^2\right)^2=0$ or $J=\bar{J}$}
In this case, Eq. (\ref{radial12222}) becomes
\begin{equation}\label{radial1222}\
z(1-z)\partial_{z}^2R(v)+\left(1-z\right)\partial_{z}R(z)+
\left[A+\frac{B}{z}+\frac{C}{1-z}\right]R(z)=0 \, ,
\end{equation}
where $A$, $B$, and $C$ are given by Eq. (\ref{A}).
Under the decomposition $R(z)=z^\alpha(1-z)^\beta K(z)$, Eq.
(\ref{radial1222}) can be written as a hypergeometric equation for $K$, Eq. (\ref{hypergeometric}),
where the coefficients $a_1$, $b_1$ and $c_1$ are given by
\begin{equation}\label{a}\
a_1= \alpha +\beta \mp \sqrt{A}~,\,
b_1=\alpha +\beta \pm \sqrt{A} ~,\,
c_1=1+2\alpha~,
\end{equation}
and the exponents $\alpha$ and
$\beta$ are
\begin{equation}
\alpha=\pm i\sqrt{B}~,\,
\beta=\frac{1}{2} \left(1\pm\sqrt{1-4 C}\right) ~.
\end{equation}

Following the same procedure detailed in the case of massive radial scalar field we obtain for $m^2/\bar{\Lambda}<0$  ($\beta_-<0$ and $\beta_+>0$)    that the field at infinity is null if  the gamma function $\Gamma(x)$ has the poles at $x=-n$ for
$n=0,1,2,...$. Then, the wave function satisfies the considered boundary
condition only upon the following additional restriction
$(c_1-a_1)|_{\beta_-}=-n$ or $(c_1-b_1)|_{\beta_-}=-n$. These conditions
determine the form of the QNFs as
\begin{equation}\label{qnm}\
\omega_1=-i\abs{\bar{\Lambda}}(r_-+r_-)(2n+1+\sqrt{1-4C})-\kappa \sqrt{-\bar{\Lambda}}\,,
\end{equation}
\begin{equation}\label{qnm}\
\omega_2=-i\abs{\bar{\Lambda}}(r_+-r_-)(2n+1+\sqrt{1-4C})+\kappa \sqrt{-\bar{\Lambda}}\,,
\end{equation}
respectively.

Now, in a similar manner to the first case, that is, using the condition that the flux vanishes at asymptotic infinity, we obtain for $\beta=\beta_+$ and $0>m^2>\bar{\Lambda}$, with $(b_1)|_{\beta_+}=-n$ or  $(a_1)|_{\beta_-}=-n$, the second set of QNFs, which yields
\begin{equation}\label{qnm}\
\omega=-i\abs{\bar{\Lambda}}(r_-+r_-)(2n+1-\sqrt{1-4C})-\kappa \sqrt{-\bar{\Lambda}}\,,
\end{equation}
\begin{equation}\label{qnm}\
\omega=-i\abs{\bar{\Lambda}}(r_+-r_-)(2n+1-\sqrt{1-4C})+\kappa \sqrt{-\bar{\Lambda}}\,,
\end{equation}
respectively.   In this case, the QNFs correspond to $\omega=\omega_{mr} \mp \kappa \sqrt{-\bar{\Lambda}}$, where $\omega_{mr}$ are the QNFs for massive radial scalar fields. So, the condition ($(-1+\bar{a})^2 J^2+4 \bar{a} \bar{\Lambda} \left(r_-^2- r_+^2\right)^2=0$ or $J=\bar{J}$) only has effect  on the real part of the QNFs. As mentioned, when $\xi=1$ ($J$=$\bar{J}$), the solution results in the BTZ black holes with a shifted cosmological constant, $\bar{\Lambda}=\Lambda- 2b^2(\lambda-1)$. Note that the QNFs have real and imaginary parts, with an imaginary part that is negative, which ensures that the propagation of scalar fields is stable in this background.



\section{Remarks and conclusions}
\label{conclusion}

In this work, we computed the QNMs of rotating three-dimensional Ho\v{r}ava AdS black holes and we analyzed the effect of the breaking of the Lorentz invariance on the QNMs.  We showed that depending on the parameters,  the lapsus function can represent a spacetime without an event horizon, a black hole geometry with one event horizon, an extremal black hole and finally a black hole with two horizons.  The QNMs have been obtained by imposing on the horizon that there
are only ingoing waves, while at infinity Dirichlet boundary conditions and
Neumann boundary conditions were imposed. We found that the propagation of the scalar field is stable in this background, since the imaginary part of the QNFs is negative. Also, we made a systematic study of the QNMs and QNFs for various ranges of parameters $J$, the angular momentum and $\xi$ the parameter that differentiates Horava gravity from General Relativity which is obtained for $\xi=1$. For various values of $J$ and $\xi$, we obtained various branches of solutions with different properties of QNMs and QNFs. In particular:

For positive inner and outer horizons $r_\pm$, the range $\xi_e < \xi < \xi_c$ gives $Re(\omega)$  null for the fundamental QNFs, and the two sectors $T_R$ and $T_L$  are well defined. $\abs{Im(\omega_1)}$ increases when the parameter $J$  increases, so according to the gauge/gravity duality, the relaxation time in order to reach the thermal equilibrium decreases. Furthermore, as the coupling constant $\xi$ increases,  $\abs{Im(\omega_1)}$ decreases and therefore  the relaxation time increases. However, for $\omega_2$, the behavior is opposite. $\abs{Im(\omega_2)}$ decreases when the parameter $J$  increases and therefore the relaxation time increases. Furthermore, $\abs{Im(\omega_2)}$ increases when $\xi$ increases; so, the relaxation time decreases.

In the range $\xi > \xi_c$ the black hole only has one horizon $r_+>0$.  For  this case  $J^2<4a^2(\xi-1)$ and the  QNFs acquire a real part, with $Re(\omega_1)=-Re(\omega_2)$,   $\abs{Re(\omega)}$ decreases when $J$ increases, and $Im(\omega_1)=Im(\omega_2)$, and it is negative. Also, when $\xi$ increases, $\abs{Im(\omega)}$ decreases. Finally, for $\xi=\xi_c$, that is, $r_-=0$, the two sectors converge. This occurs when  $J^2+4a^2(1-\xi)=0$. In this case $\abs{Im(\omega)}$ decreases when $\xi_c$ increases whereas $Re(\omega)$ is null. Therefore, the relaxation time increases. Also, we have considered different values of the constants $a$ and $b$, and $\abs{Im(\omega)}$ increases when the constant $a$ or $b$ increases; so, the relaxation time decreases.

Moreover, for the general case, that is, a massive scalar field, it was shown that the Klein-Gordon equation can be written as the Heun's differential equation, and we have studied the behavior of the QNMs numerically via the pseudospectral Chebyshev method, and mainly it was found that for $\omega_2$ and $\xi>1$, the absolute value of the imaginary part decreases as $\kappa$ increases; however, for $\xi <1$ the behavior is the opposite. The absolute value of the imaginary part increases as $\kappa$ increases.

As can be seen from the above discussion, the oscillatory and the decay modes of the QNMs have quite different behavior for the various branches, indicating that the time required for a system to reach thermal equilibrium on the boundary is  different for the various values of the parameters. An intriguing result is that if the parameter $\xi$ lies between two different critical values, then the time required for the two sectors $T_R$ and $T_L$ to reach thermal equilibrium is competing in the sense that in one sector the time is increasing while in the other sector it is decreasing. This behavior deserves further study in connection of trying to find a system on the boundary that exhibits such a behavior.

It would be interesting to extent this work to higher dimensional Horava black holes \cite{Koutsoumbas:2010pt} and calculating the QNMs and QNFs of a massive wave to study how a gravity theory in the bulk with broken Lorentz invariance affects the boundary field theory to reach thermal equilibrium.

\section*{Acknowledgments}

This work was funded by the Direcci\'on de Investigaci\'on y Desarrollo de la Universidad de La Serena (Y.V.). E.P., Y.V. and R.B would like to thank the Facultad de Ingenier\'{\i}a y Ciencias, Universidad Diego Portales for its hospitality,
where part of this work was carried out.


\end{document}